\documentclass[12pt]{article}

	\usepackage[USenglish]{babel}
	\usepackage[T1]{fontenc}
	\usepackage[utf8]{inputenc}
\usepackage{bbm}
	\usepackage{amssymb,adjustbox,amsfonts,amsmath}
	\usepackage{accents}
	\usepackage{microtype}
	\usepackage{url}
	\usepackage[comma]{natbib}
	\usepackage[singlespacing]{setspace}
	\usepackage[bottom]{footmisc}
	\usepackage{indentfirst}
\usepackage{amsmath}
\usepackage{tcolorbox}
	\usepackage{xcolor}
	\usepackage{graphicx}
	\usepackage{mathtools}

\usepackage{amsmath,amssymb}

\usepackage{amsthm}

	\usepackage{pstricks}
\usepackage{pst-all}
\usepackage{pstricks-add}
\usepackage{pst-func}
 \usepackage{pst-text}
 \usepackage{pst-plot}
\usepackage{pst-math}
\usepackage{tikz}

	\usepackage{enumerate}

	\usepackage{etoolbox}

	\let\oldciteauthor\citeauthor
	\renewcommand{\citeauthor}[1]{{\protect\NoHyper\oldciteauthor{#1}\protect\endNoHyper}}

	\let\epsilon\varepsilon
	\def\1{\mathds{1}}

	\def\de{\mathop{}\!\mathrm{d}}

	\newcommand{\g}{\ifnum\currentgrouptype=16 \;\middle|\;\else\mid\fi}
	\newenvironment{proof}[1][Proof]{\noindent\textbf{#1.} }{\ \rule{0.5em}{0.5em}\par}

	\newtheorem{theorem}{Theorem}

	\newtheorem{lemma}{Lemma}
	\newtheorem{proposition}{Proposition}

 \newtheorem*{theorem-non}{Assumption}

	\newcommand{\draftmode}[1]{
	\ifnum#1=1
		\usepackage[pass]{geometry}
		\newlength\DX
		\DX=2.5in
		\paperwidth=\dimexpr\paperwidth-\DX\relax
		\hoffset=\dimexpr\hoffset-.5\DX\relax
		\newlength\DY
		\DY=2.5in
		\paperheight=\dimexpr\paperheight-\DY\relax
		\voffset=\dimexpr\voffset-.5\DY-.5\footskip\relax
	\else
		\usepackage[vcentering,nohead,margin=1in]{geometry}
	\fi
	}

	\newcommand{\showlabels}[1]{
	\ifnum#1=1
		\usepackage[inline]{showlabels}
	\fi
	}

	\newcommand{\showcomments}[1]{
	\ifnum#1=1
		\definecolor{NLcolor}{RGB}{30, 126, 56}
		\newcommand{\NL}[1]{{\itshape\textbf{[NL:} {\color{NLcolor} #1}\textbf{]}}}
	\else
		\newcommand{\NL}[1]{}
	\fi
	}

	\let\originalleft\left
	\let\originalright\right
	\renewcommand{\left}{\mathopen{}\mathclose\bgroup\originalleft}
	\renewcommand{\right}{\aftergroup\egroup\originalright}

	\makeatletter
	\newcommand{\Matrix}[1]
	{\begin{bmatrix}
	  \Matrix@r #1;\@bye;\Matrix@r
	 \end{bmatrix}}
	\def\Matrix@r #1;{\@bye #1\Matrix@z\@bye\Matrix@s #1,\@bye, }
	\def\Matrix@s #1,{#1\Matrix@t }
	\def\Matrix@t #1,{\@bye #1\Matrix@y\@bye\@firstofone {&#1}\Matrix@t}
	\def\Matrix@y #1\Matrix@t{\\ \Matrix@r }
	\def\Matrix@z #1\Matrix@r {}
	\def\@bye  #1\@bye {}
	\makeatother

	\def\<#1>{\Matrix{#1}}

\newcommand{\Rbar}{\overline{\mathbf{R}}}
\newcommand{\R}{\mathbf{R}}

		\newcommand{\bep}{\bar{\epsilon}}

\draftmode{0}
\showlabels{0}
\showcomments{0}
\usepackage{datetime}
\newdateformat{datestyle}{\monthname[\THEMONTH], \THEYEAR}

 \title{Separating the Wheat from the Chaff\thanks{We thank J\'{e}r\^{o}me Renault, Fabien Gensbittel and David Levine for enlightening discussions on existence, Kourosh Khounsari for excellent research assistance, and several seminar audiences for helpful feedback.}}

\author{
\small
\begin{tabular}{cc}
	Johannes H\"{o}rner & Paula Onuchic\\
	CNRS (TSE) & London School of Economics
\end{tabular}
}
\date{\small November 2026}

\begin{document}
\maketitle

\begin{abstract}
We study a reputational cheap-talk environment in which a judge chooses between two speakers of unknown expertise, an expert and a quack.  Both speakers seek to be selected, while the judge wishes to identify the expert. The best equilibrium for the judge features truth-telling by the expert, yet the judge favors speakers with more extreme messages. This bias toward extremism does not induce exaggeration, but instead sustains truthful communication. We characterize the quack's equilibrium messaging strategy, relating the degree to which it panders to the judge's prior to the state distribution, the speakers' initial reputations, and the judge's private knowledge of the state. We interpret these results in the context of two applications, the post-training of AI chatbots through human feedback and media competition.

\end{abstract}

\section{Introduction}
This paper studies a three-player reputational cheap-talk game. A judge must select one of two speakers, whom she deems to be an expert  informed about a relevant state. One of the speakers is indeed an expert, while the other is an uninformed quack. The speakers are not neutral conveyors of their (lack of) knowledge, but instead motivated by the prospect of being recognized and selected by the judge as the expert. In such an environment, we ask three questions:
Should genuinely informed experts be expected to report the truth? How should a quack strategically tailor his claims in order to appear knowledgeable? And how should the judge, equipped with only imperfect information of her own, decide whom to trust?

We study equilibrium communication in this context, showing that reputational competition can sustain truthful communication by experts, characterizing patterns of speech made by quacks, and describing how audiences learn from these competing sources. 

Our leading application of this model is to the comparative evaluation of AI chatbots. In that setting, we view the judge as a human evaluator who is presented with two perhaps conflicting answers to the same query and must select which one to reward. Such comparisons can be used to post-train the systems or to determine which system is ultimately deployed. One chatbot has access to the information needed to answer the query while the other is ungrounded but can nevertheless produce a plausible answer. Neither chatbot is adversarial to the user or intentionally deceptive; they are solely rewarded for being perceived as grounded. Applied to this setting, our analysis characterizes how reputational incentives shape truthful reporting by grounded AI systems, equilibrium patterns of hallucinations produced by ungrounded  systems, and the user's ability to identify the informed system and learn from its reports.

Media competition provides a second application. Here, the judge is an audience who sees accounts on a particular theme provided by multiple media sources, assesses which source appears better informed or more credible, and chooses which one to rely on. Each competing media source seeks only to be chosen by the audience. The model therefore offers a theory of learning in media environments where sources are plentiful, their reliability is uncertain, and each has reputational incentives to appear well informed.

Section \ref{sec:benchmark} introduces our benchmark model, which takes a deliberately stark view of expertise. One speaker, the expert, perfectly observes a relevant state, while the other, the quack, is completely uninformed. The judge lies between these extremes: she does not observe the state, but receives an imperfect private signal about it. The benchmark  adopts simple distributional assumptions, taking both the state and the noise in the judge's signal to be uniformly distributed. 

The speakers simultaneously send cheap-talk messages. The judge then observes both messages together with her private signal and selects one of the speakers. The judge's goal is to select the expert speaker, while each speaker wishes only to be selected by the judge.  Communication about the state is therefore purely instrumental: speakers care about what their reports imply for their perceived expertise, and the judge uses those reports to distinguish informed communication from uninformed imitation.

Our main result, Theorem \ref{thm:best}, characterizes the best equilibrium for the judge, that is, the one that maximizes the \textit{ex ante} probability that the judge selects the true expert. We show that, in this equilibrium, (1) the expert reports the state truthfully, (2) rather than concentrating on messages that are more likely to appear plausible to the judge, the quack exactly replicates the expert's marginal message distribution, which in turn corresponds to the state distribution, and (3) the judge uses a selection rule that favors the more extreme speaker,  whose message is \textit{ex ante} less likely to appear plausible in light of the judge's private information.

In this equilibrium, both speakers use the same marginal message distribution, but the expert's message is correlated with the  state. Knowing this correlation structure, the judge can infer the speakers' expertise by assessing whether their speech is consistent with her own private signal. If only one of the two observed messages is consistent, the judge learns the identity of the expert, as well as the state. When instead both messages are consistent, the judge does not learn the expert's identity and remains indifferent between selecting either speaker.
The selection rule used by the judge in such indifference cases is a key equilibrium object: Our construction shows that, by favoring the selection of the speaker who used the more extreme message, the judge offsets the quack's incentive to pander to moderate messages  and ensures that truth-telling is an optimal strategy for the expert.

In Section \ref{sec:features}, we describe how the judge learns about the state in the truth-telling equilibrium described in Theorem \ref{thm:best}, and how this learning depends on the precision of the judge's own private signal. One central feature is that ``information begets information:'' a judge with a more precise private signal is more likely to separate the expert from the quack and further learn about the state. Through this mechanism, communication from speakers of unknown expertise can amplify informational differences between judges who begin with private signals of different precisions. Whether it does so, however, depends on how learning is measured: Proposition \ref{pr:2} shows that judges with more precise private signals are more likely to learn the state with certainty, whereas Proposition \ref{pr:var} shows that less precise judges obtain a larger reduction in expected posterior variance.

The benchmark model in Section \ref{sec:benchmark} is deliberately stark, both in its simplifying distributional assumptions and in taking an extreme view of expertise. In Sections \ref{sec:state} and \ref{sec:identity} we consider extensions of our benchmark environment in which the distribution of the state is non-uniform or the judge has prior information about the identity of the expert. Further, Section \ref{sec:generalmodel} introduces a model of ordered expertise, where expert and quack are labels that refer to the speakers' ordered knowledge of the state. This model also makes more general distributional assumptions on the state and the judge's signal. In all extensions, the judge's best equilibrium continues to feature truth-telling by the expert. 

What changes is the behavior of the quack, whose messaging strategy can pander to more moderate messages, corresponding to states the judge is more likely to deem plausible. The extent and form of the quack's pandering depend on the extension. In Sections \ref{sec:state} and \ref{sec:identity}, non-uniformity of the state distribution and of speakers' initial reputations lead to pandering that operates on the extensive margin: the quack no longer makes reports that are ``too extreme.'' Instead in Section \ref{sec:generalmodel}, under non-uniformity of the judge's information, pandering occurs on the intensive margin: all reports remain in use, but their relative frequencies shift towards more moderate messages.

Across the benchmark and all extensions, our results also establish a strong optimality property of truthful communication. Fixing the expert's strategy at truthful reporting, the interaction between the judge and the less-informed speaker has a zero-sum structure: the judge chooses the correct expert exactly when the quack is not selected. We leverage this zero-sum property of the truth-telling game to show that the truthful equilibrium attains the highest \textit{ex ante} payoff the judge could achieve even if she could commit in advance to a communication mechanism. Thus, neither commitment nor a more elaborate form of debate can improve her ability to identify the expert speaker.

Section \ref{sec:applications} discusses our main applications. In the AI application, our main result shows that the comparative evaluation of AI chatbots by a human evaluator is sufficient to sustain truthful reporting by the grounded chatbot. The equilibrium also yields a theory of AI hallucinations. Because the grounded chatbot reports truthfully, all hallucinations in the model originate with the ungrounded chatbot. The model predicts not only their source but also their shape: In the benchmark, the distribution of hallucinations replicates the marginal distribution of truthful answers. In our extensions, they can shift towards safer and more sycophantic answers. Our results relating the structure of hallucinations to the distribution of correct answers, the structure of the user's private information, and the systems' initial reputations provide empirically testable predictions of the theory.

Our application to media competition highlights how exposure to the same potentially unreliable media sources can amplify informational inequalities present in a population (corresponding to a set of judges with signals of different accuracies, in our model). We discuss this theory of information inequality in contrast with other ``partisan echo chamber'' theories of media-driven polarization.

\subsection{Related Literature}
Our paper contributes to the vast literature on cheap talk, starting with Crawford and Sobel (1982) and Green and Stokey (2007), and especially that on reputational cheap talk. Our paper is more closely related to models in which senders are concerned about their reputation for expertise, such as Scharfstein and Stein (1990), Prendergast (1993), Prendergast and Stole (1996), Ottaviani and S\o rensen (2001, 2006a, 2006b), and Levy (2004).\footnote{Other models of reputational cheap talk -- such as Morris (2001), Olszewski (2004), or Brandenburger and Polak (1996) --- consider senders who wish to be perceived as unbiased, honest, or as making high-quality decisions. See also Friedman (1998), in which the speaker actively decides whether to become an expert, a choice that depends on the degree of their bias.}
A common theme of these papers is that the sender slants their reports toward the receiver's prior beliefs (``reputational herding'').\footnote{Reputational concerns can also generate anti-pandering incentives, as in Prendergast and Stole (1996) or Levy (2004). In those papers, anti-pandering takes the form of a distortion by the reputationally motivated agent away from the receiver's prior. In a one-speaker version of our model in Online Appendix OA.6, we similarly show that equilibrium can feature ``anti-pandering'' by the quack.}

Ottaviani and S\o rensen (2006b) show that, in a generic informational environment, the speaker's incentives to pander to the receiver's prior prevent the existence of truth-telling equilibria.
In our context, truth-telling by the expert is a robust feature of equilibrium, and non-pandering by the quack is also a non-degenerate equilibrium outcome.
A key distinction between these settings is that our judge is an active decision-maker. In Ottaviani and S\o rensen (2006b), the sender's payoff is directly tied to the receiver's posterior belief about their expertise. In our model, by contrast, the judge forms a posterior as to which speaker is most likely to be the expert, but she still actively selects one of the two speakers. This additional discretion is consequential.

Our analysis in Section \ref{sec:arg1} shows that if, whenever indifferent, the judge selects between the two speakers uniformly, then our model reproduces the pandering force in Ottaviani and S\o rensen (2006b). Allowing the judge instead to condition her tie-breaking on the realized messages overturns this incentive.
Thus our analysis identifies active selection by the receiver as a force that can undo the reputational-herding incentives emphasized in  previous work.\footnote{We view the contrast between  reputational rewards based directly on posterior beliefs or those mediated by an active decision by the receiver as representing different economic settings. Direct belief-based utility is fitting when agents value esteem or being regarded as competent \textit{per se}. In many applications, however, beliefs about expertise matter because they affect a later choice: an audience chooses which media source to follow, an employer chooses whom to hire or promote, or an evaluator chooses which AI system to reward or deploy. In such settings, treating posterior reputation itself as the payoff can suppress economically relevant discretion in how equally credible agents are rewarded. Accordingly, our analysis then implies that the (non-)existence of truth-telling equilibria may itself depend on the economic setting.}

In our model, the receiver consults two speakers, rather than a single sender. Some previous literature, including Krishna and Morgan (2001) and Battaglini (2002), also considers models of cheap talk with multiple senders. Our paper relates also to work on the optimal structure of debates. For example, Ottaviani and S\o rensen (2001)
show that changing the order of speech can improve information aggregation, and Glazer and Rubinstein (2001) show that sequential communication can be used to economize on information transmission costs.\footnote{Glazer and Rubinstein (1998) study more generally how experts' motives affect the design of mechanisms for eliciting information. In particular, they show that adding a motive to have one's recommendation accepted can change what outcomes are implementable.} In our setting, simultaneous communication achieves the best payoff to the judge. 

Our main application considers post-training via the comparative evaluation of AI chatbots. Most of the related literature in economics focuses on the AI \emph{alignment} problem --- see, for example, Chen, Ghersengorin, and Petersen (2024),  Fudenberg and Liang (2026), and Dworczak and Smolin (2026) --- discussing how humans can identify, take advice from, and delegate to, AI tools whose incentives might be misaligned with (or even adversarial to) those of the user.\footnote{Among these papers, Dworczak and Smolin (2026) is closest to ours. As in our model, a human decision-maker has private information about the state and must evaluate information provided by an AI whose reliability is uncertain. They show that optimal reliance on AI takes the form of a ``trust region:'' sufficiently extreme reports are clipped, limiting the influence of reports that could be especially harmful.} In our model, the AI tools are not intrinsically aligned or adversarial, but simply seek the ``approval'' of the human user. We show that the comparative evaluation of two chatbots can produce the incentives for them to act as if aligned with the human's interests, in the sense that the grounded bot truthfully reports the state.


Our paper also provides a theory of sycophantic hallucinations and their connection to  post-training. This connection has been recognized empirically in the computer-science literature.\footnote{Sharma \textit{et al.} (2024) show that human preference data rewards responses that match users' beliefs, and that optimizing against preference models can sometimes increase sycophancy and sacrifice truthfulness. Huang \textit{et al.} (2025), in their survey of LLM hallucinations, accordingly identify reinforcement learning from human feedback as one potential source of hallucination and describe sycophancy as a form of belief misalignment in which models seek human approval at the expense of truthfulness.} Our contribution is to formalize it through the incentives generated by comparative evaluation. The theory yields two additional implications. First, the prediction that hallucinations originate with the ungrounded system, while the grounded system reports truthfully. Second, it relates the distribution of hallucinations --- and in particular their degree of sycophancy --- to features of the evaluation environment, generating potentially testable predictions about the shape of model errors.

Our second application is to media competition. Gentzkow and Shapiro (2006) is particularly close to our work in that motivation: they interpret media companies as speakers who wish to develop a reputation for expertise. They show that media companies pander to the prior held by their audiences, especially so if their audience has little access to other information sources. One implication is that information polarization or inequality can arise because different media outlets cater to pre-existing heterogeneous priors across a population. Perego and Yuksel (2022) show that media competition increases disagreement by inducing specialization in issues on which audiences' preferences differ.\footnote{Related experimental work documents both audiences' tendency to view congenial information as more credible and speakers' incentives to cater to such beliefs. Thaler (2024) finds that subjects judge politically congenial news as more likely to come from a truthful source. Thaler (2025) shows that incentives to be perceived as truthful can induce senders to tailor false reports to receivers' politically motivated beliefs. Closest to our source-selection problem, Alysandratos \textit{et al.} (2023) find that many subjects fail to identify an expert when comparing expert and populist advice.} Our paper proposes an alternative explanation for the coexistence of widespread access to information and growing information inequality, which does not appeal to media companies pandering to their audience.\footnote{Relatedly, Andreoni and Mylovanov (2012) show how two-dimensional uncertainty --- in our case, about the state and about the speakers' expertise --- might lead to divergence of posterior beliefs along a one-dimensional subspace.}

\section{Benchmark Environment}\label{sec:benchmark}
There are three players, two speakers ($i=1,2$) and a judge. There is an unknown state of the world $\omega$ in $[-1,1]$, which is distributed uniformly. The speakers differ in terms of their knowledge of the state: one speaker is an expert, who perfectly observes the state, and one speaker is a quack, who knows nothing about the state. The speakers know their own (and each other's) identities, but the judge believes each speaker is \textit{ex ante} equally likely to be the expert or the quack.\footnote{In line with the implementation theory literature, \textit{e.g.}, Maskin (1999), we assume that the speakers’ types are common knowledge among them, though unknown to the judge, therefore focusing attention on the receiver’s inference problem rather than on uncertainty among senders.}

Each speaker $i$ sends a message $m_i$ from the set $\mathcal{M}=[-1,1]$ to the judge. The speakers send their messages simultaneously. The judge then selects one of them. The judge's objective is to select the correct speaker --- the expert --- and each of the speakers wishes to be selected.
Before selecting a speaker, the judge sees both speakers' messages, and also receives a private noisy signal about the state, $s=\omega+\epsilon$, where $\epsilon\sim\mathcal{U}[-\bar{\epsilon},\bar{\epsilon}]$. The parameter $\bar{\epsilon}$ in $(0,1]$ measures the precision of the judge's private signal.
The judge receives a payoff of $1$ if she selects the correct speaker, and $0$ otherwise; each speaker receives a payoff of $1$ if they are selected, and $0$ otherwise.

Our equilibrium concept is Perfect Bayesian Equilibrium: the expert and quack, respectively, use messaging strategies $F_E$ in $\Delta ([-1,1]\times \mathcal{M})$ --- a joint law with uniform marginal on $[-1,1]$ --- and $F_Q$ in $\Delta\mathcal{M}$ that maximize the probability of being selected; and the judge uses a measurable selection rule $\phi:\mathcal{M}\times\mathcal{M}\times[-1-\bar{\epsilon},1+\bar{\epsilon}]\rightarrow\Delta\{1,2\}$, which is supported only on the speaker (or speakers) that is most likely to be the expert, given the judge's Bayes-consistent beliefs after seeing the realization of her private signal and the speakers' pair of messages.

\paragraph{Modeling Choices.} Our benchmark environment makes restrictive uniformity assumptions about the distribution  of the state, of speakers' expertise types, and of the judge's private signal errors. After establishing our benchmark results, we discuss their extensions to settings in which we generalize each of these uniformity assumptions --- Sections \ref{sec:state} and \ref{sec:identity} consider a non-uniform state distribution and non-uniform prior over the speakers' expertise types, respectively; Section \ref{sec:generalmodel} discusses a general model of ordered expertise, in which the expert might be imperfectly informed, the quack might have some information about the state, and the judge's signal error is not uniform.

We now highlight the interpretation of our benchmark modeling choices.

\begin{enumerate}
\item \textbf{Two speakers, with anti-correlated types}. In our environment, exactly one speaker is an expert and the other a quack --- in contrast, for example, with a setting in which speakers' expertise types are drawn independently.\footnote{If speakers' types are \textit{i.i.d.}, the other speaker's report can be included among the judge's observations when assessing either speaker. The selection decision remains comparative. Section OA.6 of the online appendix studies a single speaker facing a fixed outside option.} The judge's problem is therefore comparative, that of distinguishing informed communication from uninformed imitation. Accordingly, a speaker’s credibility is relative; each speaker’s persuasiveness depends on how their message compares to the other's. Our model therefore speaks primarily to settings in which sources with different expertise levels \emph{compete} for an audience's trust --- for example, competitive debates, conflicting court testimony, competing media sources, or side-by-side testing of AI chatbots when systems are grounded in distinct data or tools.

\item \textbf{Judge's private signal.} The assumption that noise in the judge's signal is uniform yields the interpretation of the judge's information as domain heuristics: she has access to
elementary facts about the issue at hand that can be used to spot communications that cannot be consistent with reality. Say, she might know that a specific concept (\textit{e.g.}, zero) was not used in Roman numerals, allowing her to detect a pseudo-historian pretending to be an expert on Roman history. The judge's truth-assessment technology is constrained to  examining whether or not the claims made by each speaker contradict their own `meta-data.' Such binary tests allow the judge to detect a cheat conclusively if he fails, but are uninformative when he passes through luck.
The judge's accuracy is determined by the support parameter $\bar{\epsilon}$. More accuracy (a lower $\bar{\epsilon}$) might reflect a higher degree of education of the judge, her knowing more ``facts,'' for example.

The structure of the judge's signal also means that ``moderate'' states --- those closer to $0$ --- are more likely to be within the range that the judge deems plausible than ``extreme'' states, those closer to $-1$ or $1$.\footnote{In our generalization to other possible distributions of the state and of the judge's signal, we maintain the feature that states belong to a bounded interval, and that the judge's signal is centered around the true state (a location experiment, in the sense of Lehmann, 1988). As in the baseline, the judge is therefore likely to deem moderate states as more plausible.} Accordingly, our use of ``extreme'' to describe states and, by extension, speech refers to this lower \textit{ex ante} likelihood of being deemed plausible; it carries no connotation of political extremism or inflammatory rhetoric.

\item \textbf{Judge actively selects a speaker.} In our game, the judge actively selects a speaker after hearing both messages and her own private signal, and the speakers' payoffs depend only on the judge's selection. This is in contrast with previous models of reputational cheap talk --- for example, Ottaviani and S\o rensen (2006a, 2006b) --- in which the speaker's payoff is a direct function of the observer's posterior belief about their type. Finally, we assume throughout that the judge \textit{must} pick a speaker. She cannot abstain, and transfers are not permitted, ruling out resolutions to the judge's conundrum along the lines of King Solomon's dilemma (\textit{e.g.}, Glazer and Ma, 1989).\footnote{If abstention were possible and the judge could commit to a selection rule, then she would be able to  perfectly extract information about the expert's true identity, since it is known by both speakers. }
\end{enumerate}

\subsection{Main Result}\label{sec:ch}
The reputational cheap talk model described above has many equilibria. As usual, it has a babbling equilibrium, in which both the expert and the quack's messaging strategies are independent of the state and the judge selects a speaker at random. There are also various equilibria with information transmission.\footnote{Such as one in which the expert reports the sign of the observed state ($+$ or $-$), the quack randomizes between $+$ and $-$, and the judge selects the sign that matches her observed signal, if only one of the messages does, and randomly selects a speaker otherwise.} Our main result characterizes the \emph{best equilibrium for the judge}, which maximizes the judge's \textit{ex ante} probability of selecting the expert.

\begin{theorem}\label{thm:best}
There exists an equilibrium in which the expert reports the true state $\omega$ and the quack's messaging strategy replicates the prior state distribution, $\mathcal{U}[-1,1]$.  Its payoff to the judge is maximal among all equilibria, including those of alternative communication mechanisms. These strategies can be supported in equilibrium by different selection strategies $\phi$ used by the judge to reward speakers who send extreme messages.
\end{theorem}

We prove this result in three steps: (1) We characterize selection strategies $\phi$ for the judge that make the quack indifferent between all messages, conditional on the expert being truthful; (2) A truth-telling result verifies the expert's incentives; (3) A general optimality proposition shows that the resulting equilibrium attains the commitment bound.

\subsubsection{Quack's Indifference Condition}\label{sec:arg1}
For this argument, suppose the expert always sends the truthful message, so that $F_E(\cdot|\omega)=\delta_\omega$, and the quack randomizes his message according to the prior state distribution, so that $F_Q=\mathcal{U}[-1,1]$.\footnote{Recall that $F_E$  and $F_Q$ are the distributions  used by the expert and the quack, respectively; here and in what follows, $F_E(\cdot|\omega)$ denotes a version of the conditional distribution used by the expert, given the state $\omega$. For convenience, we often refer to the cdf as the distribution.} Given these strategies, and the judge's private signal $s$, there are two circumstances the judge might face:

\begin{enumerate}
\item Exactly one message --- the expert's message --- is \emph{consistent} with the judge's signal $s$, where we say a message $m$ is consistent with signal $s$ if $|m-s|\leqslant\bar{\epsilon}$.

\item Both messages $m_1$ and $m_2$ are consistent with the judge's signal.
\end{enumerate}

Note that a circumstance in which both messages are inconsistent cannot arise, since the expert is assumed to always reveal the state truthfully.

If the first circumstance occurs, then the judge perfectly identifies the expert to be the speaker who sent the consistent message. She then optimally selects that speaker. In the second circumstance, the two identity assignments instead have the same likelihood. Up to their common identity prior, the likelihoods of $(m_1,m_2,s)$ are
\[
 \frac12\frac12\frac1{2\bep}\mathbf 1\{|s-m_2|\leqslant\bep\}
 \quad\text{and}\quad
 \frac12\frac12\frac1{2\bep}\mathbf 1\{|s-m_1|\leqslant\bep\},
\]
respectively if speaker 2 or speaker 1 is the expert. These values coincide when both messages are consistent, so the judge can then optimally select either speaker.

\paragraph{Uniform Tie-Breaking.} To understand the construction, we first argue that a uniform tie-breaking rule \emph{cannot} satisfy the quack's indifference condition. Suppose momentarily that, conditional on both messages being consistent, the judge selects each of the speakers with equal probability. In this case, the quack's objective is to maximize the \textit{ex ante} probability that his message is consistent with the judge's drawn signal, that is, he chooses a message $m$ so as to maximize $\mathbf{P}[|m-s|\leqslant\bar{\epsilon}]$.

The judge's signal, being a sum of two  independent uniform random variables, follows a trapezoidal distribution.\footnote{Specifically, for $|s|\le 1+\bep$,
\[\mathbf{P}(s)=\begin{cases}
\frac{1+\bar{\epsilon}}{4\bar{\epsilon}}+\frac{s}{4\bar{\epsilon}}\text{, if }s\leqslant-(1-\bar{\epsilon}),\\
\frac{1}{2}\text{, if }s\in[-(1-\bar{\epsilon}),(1-\bar{\epsilon})],\\
\frac{1+\bar{\epsilon}}{4\bar{\epsilon}}-\frac{s}{4\bar{\epsilon}}\text{, if }s\geqslant1-\bar{\epsilon}.
\end{cases}\]}
Consequently, the objective $\mathbf{P}[|m-s|\leqslant\bar{\epsilon}]$ is maximized by the moderate messages satisfying $|m| \le (1-2\bar{\epsilon})_+$, which are therefore strictly preferred by the quack to all messages outside of this range. In particular, the quack is not willing to randomize over the entire support of possible messages, as we initially conjectured.

\paragraph{Rewarding Extreme Messages.} From the description above, it is clear that, to make the quack indifferent between  every message on $[-1,1]$, the judge must use a selection rule that rewards extreme messages whenever they happen to be consistent with the judge's signal. Such rewarding compensates for the fact that these extreme messages are \textit{ex ante} less likely to be consistent.\footnote{There are many such rules that perfectly balance the quack's incentives. Indeed, the degrees of freedom in the judge's tie-breaking selection function --- which may depend on $m_1$, $m_2$, and $s$ --- are more than sufficient to satisfy the indifference requirements for the quack. 

Our analysis characterizes a ``max'' selection rule that provides quack indifference. In Online Appendix OA.1, we alternatively define a class of ``min'' selection rules, which define the judge's tie-breaking as a function of the least extreme message used by the speakers. Analogously to the ``max'' class, we constructively show that there is a unique ``min'' selection rule that yields the quack's indifference condition.} 

More formally, we denote by $\Phi^{ind}$ the set of selection rules for the judge that (i)  select the consistent speaker when only one message is consistent with her drawn signal and (ii) make the quack indifferent between all messages, if the expert is truth-telling.
We now construct two such selection rules, which we denote as the ``max rule'' and the ``min rule.''

\paragraph{``Max'' Selection Rule.} We define the class of ``max'' selection rules as follows: a selection rule $\phi$ belongs to $\Phi^{max}$ if, when both messages are consistent --- so that $|m_1-s|, |m_2-s|\leqslant \bar{\varepsilon}$ --- and message $i$ is the most extreme --- $|m_i|>|m_{-i}|$,
\begin{align*}
\phi(m_1,m_2,s)&=\phi_{\max}(\max\{|m_1|,|m_2|\})\delta_i\\
&+(1-\phi_{\max}(\max\{|m_1|,|m_2|\}))\delta_{-i},
\end{align*}
for some continuous $\phi_{\max}:[0,1]\to[0,1]$. When $|m_1|=|m_2|$, the judge selects each speaker with probability $1/2$.

According to a ``max'' selection rule, conditional on both messages being consistent, the judge selects the speaker whose message has the larger absolute value (the more extreme message) with a probability that depends only on the magnitude of that extreme message, so that $\phi_{\max}(\max\{|m_1|,|m_2|\})$ is the probability that the extreme speaker is selected.

\begin{lemma}\label{lem:1}
There exists a unique continuous function $\phi_{\max}:[0,1]\to[0,1]$ that can be completed to a selection rule in $\Phi^{ind}\cap\Phi^{max}$. The function $\phi_{max}$ is strictly increasing over $[0,1]$ and  onto $[1/2,1-\bep/3]$.
\end{lemma}

   \begin{figure}
\centering
\includegraphics[width=0.6\textwidth]{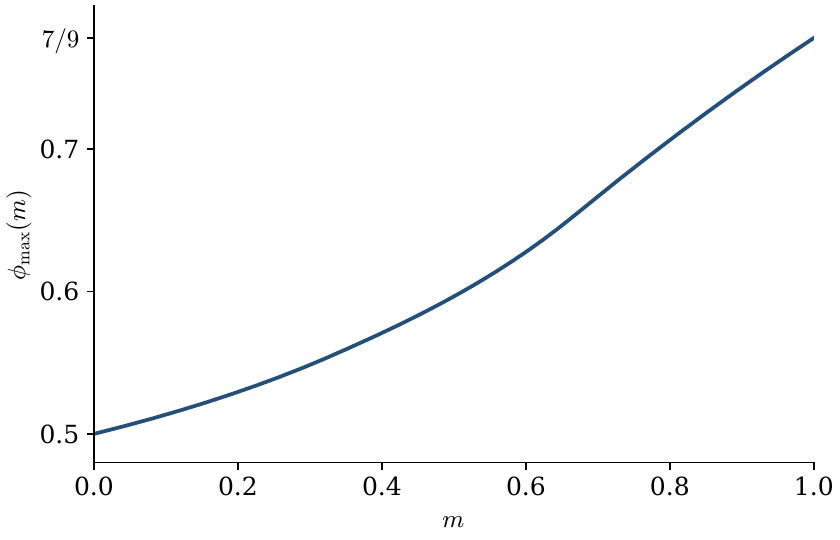}
\caption{The max rule for $\bep=2/3$.}
\label{fig:maxim}
\end{figure}
Figure \ref{fig:maxim} illustrates this selection rule, for a particular value of $\bar{\epsilon}$.  
The map $\phi_{\max}$ is strictly increasing throughout the support of possible messages, despite all messages $|m| \le 1-2\bep$ being consistent with the same probability. This is because the selection rule is solved for globally: very extreme messages are often not consistent, and  must thus be selected with higher probability when consistent. In turn, this means that slightly lower messages ``lose out'' on average to such messages, and the loss must be compensated for by above-average odds against even lower messages. This logic ripples out down to the least extreme message, $0$.

The decision rule in Lemma \ref{lem:1} is defined recursively over the intervals of messages $[1-2\bep,1]$, $[1-4\bep,1-2\bep]$, and so on, through the quack's equal-payoff condition. Consider a small value of $\bar{\epsilon}$, such that these initial intervals only involve messages larger than $2\bar{\epsilon}$.\footnote{See Appendix \ref{sec:prolem1} for details on all intervals and values of $\bar{\epsilon}$.}
Let $\Pi$ denote the quack's expected payoff, which we conjecture is constant across all messages. Consider each message $m$ in $[1-2\bep,1]$ (and suppose $1-4\bep>2\bep$). The equality condition is given by
\begin{align}
\hspace*{-.5cm}\Pi&=\phi_{max}(m) \int_{m-2\bep}^m \frac{1}{2}\left(1-\frac{m-m'}{2\bep}\right)\de m'\label{eq:int}\\[1em]
&+\int_{m}^1 \frac{1}{2}\left(1-\frac{m'-m}{2\bep}\right)\left(1-\phi_{max}(m')\right)\de m'.\nonumber
 \end{align}

To understand this expression, consider the first term on the right-hand side. If the quack's competitor, the expert, uses a message that is less extreme than that used by the quack --- $m'<m$, which happens with density $1/2$ --- then the quack is chosen with probability $\phi_{\max}(m)$ in case his message is consistent. Since $m'$ is the true state --- remember our equilibrium conjecture that the expert reports the state truthfully --- the judge's signal is uniformly distributed over the interval $[m'-\bep,m'+\bep]$; and so the quack's message is consistent if, and only if, $s \ge m-\bep$, which occurs with probability $1-\frac{m-m'}{2\bep}$. The second term reflects the alternative possibility that the expert uses a more extreme message than the quack --- $m'>m$, which occurs with density $1/2$ --- in which case the probability with which the quack is chosen is $1-\phi_{\max}(m')$, if his message is consistent.

Combining these two events, we have an integral equation for $\phi_{max}$, which involves the parameter $\Pi$, the quack's expected payoff. The equilibrium value of $\Pi$ can be inferred from a separate condition: the quack's expected payoff is the probability that the judge makes a mistake, which must equal $1/2$ times the probability that both messages are consistent. (Remember that the judge is indifferent between selecting either speaker in that case, and so equally likely to be right or wrong in her choice.) Consequently, given that the quack's message $m$ is distributed uniformly over $[-1,1]$, and the judge's signal $s$ is independently drawn according to $\mathbf{P}$,
\begin{equation}\label{eq:payoff}
\Pi=\frac{1}{2}\int_{-1}^{1} \frac{1}{2} \int_{m-\bep}^{m+\bep}\mathbf{P}(s)\de s\de m=\frac{1}{2}\left(\bep-\frac{\bep^2}{3}\right).
\end{equation}
Setting $m=1$ in equation (\ref{eq:int}), and using the expression for $\Pi$ in (\ref{eq:payoff}) immediately yields $\phi_{\max}(1)=1-\frac{\bep}{3}$. More generally, equation (\ref{eq:int}) admits as unique solution, for $m$ in $[1-2\bep,1]$,
 \begin{equation}\label{eq:phi0}
\phi_{max}(m)=\phi_0(m):= 1-\frac{\bep }{3}  e^{ \frac{1-m}{2 \bep }} \left(\sin \left(\frac{1-m}{2
  \bep }\right)+\cos \left(\frac{1-m}{2 \bep }\right)\right).
 \end{equation}
This is the first step of the induction. Next, for $m$ in $[1-4\bep,1-2\bep]$, assuming $1-6\bep>2\bep$, the equality condition is
{\footnotesize\begin{align*}
\Pi=\phi_{max}(m)&\int_{m-2\bep}^m \frac{1}{2}\left(1-\frac{m-m'}{2\bep}\right)\de m'+\int_{m}^{1-2\bep}\frac{1}{2} \left(1-\frac{m'-m}{2\bep}\right)\left(1-\phi_{max}(m')\right)\de m'\\[1em]
&+\int_{1-2\bep}^{m+2\bep}\frac{1}{2} \left(1-\frac{m'-m}{2\bep}\right)\left(1-\phi_0(m')\right)\de m',\end{align*}} 
an equation that can be turned into a linear differential equation with constant coefficients for the second antiderivative of $\phi_{max}$.  The construction proceeds recursively, adjusting the integral equation  once $m<\bep$.\footnote{The intervals for messages lower than $\bep$ introduce the additional complication that the judge might have to decide between such a consistent message and a more extreme consistent message whose value is negative. The full construction, which we just sketched, is available in Appendix \ref{sec:prolem1}.}

\begin{figure}[t!]
\centering
\includegraphics[width=0.7\textwidth]{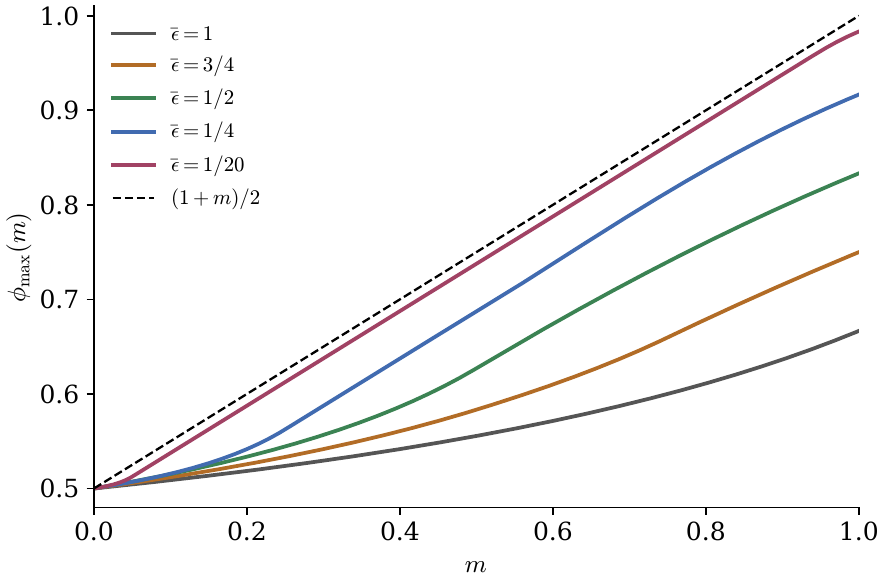}
\caption{The max rule for $\bep=1,3/4,1/2,1/4,1/20$.
The dashed line is $(1+m)/2$.}
\label{fig:myelement}
\end{figure}

\paragraph{Signal Accuracy.}
The quality of the judge's information does not affect the quack's
strategy in the equilibrium of Theorem \ref{thm:best}.
It does affect the selection rule. Figure \ref{fig:myelement} illustrates
the max rule for several noise levels. These functions need not be
ordered at every message. Their endpoint difference is
\[
\phi_{\max}(1)-\phi_{\max}(0)=\frac12-\frac{\bep}{3},
\]
which increases with signal accuracy.

More generally, quack indifference implies that the probability of
selection, conditional on consistency, is
\begin{equation}\label{eq:p2}
\zeta(m)=
\frac{\bep/2-\bep^2/6}
{\displaystyle\bep-
\frac{(m+2\bep-1)_+^2+(-m+2\bep-1)_+^2}{8\bep}}.
\end{equation}
Here $x_+=\max\{x,0\}$. The denominator is the probability that the
quack's report $m$ is consistent. Thus $\zeta$ is even and nondecreasing
in $|m|$. There is no general single-crossing comparison as $\bep$
varies. For $\bep\le1/2$, the endpoint difference is
$\zeta(1)-\zeta(0)=1/2-\bep/6$.

\subsubsection{Expert Incentives}\label{sec:arg2}
The second step in the proof of Theorem \ref{thm:best} considers the expert's strategy.

\begin{lemma}\label{lem:3}
If the quack uses messaging strategy $F_Q=\mathcal{U}[-1,1]$, and the judge uses some selection rule $\phi$ in $\Phi^{ind}$, then the truth-telling messaging strategy $F_E(\cdot|\omega)=\delta_\omega$ is optimal for the expert. Under the max or min selection rule, it is uniquely optimal.
\end{lemma}

Intuitively, in our conjectured equilibrium, the quack and expert face similar problems: from his uninformed perspective, the quack understands that the expert uses a uniform messaging strategy (this is the marginal distribution of the expert's messages); and similarly, the expert faces an opponent that uses a uniform messaging strategy.

The quack's indifference condition guarantees that, when facing such an opponent, there is no message that is better in expectation, without knowledge of the state. Of course, the expert's advantage is that he does have knowledge of the state; we show in Lemma \ref{lem:3} that truth-telling is optimal because a false report can only lose opportunities for selection when both reports are consistent. Under the max rule, this loss is positive for every false report.

\paragraph{Off-Path Selection Rule.} When considering a possible deviation from truth-telling, the expert places positive probability on the realization of report profiles with two inconsistent messages. The selection rules in $\Phi^{ind}$ do not specify the judge's strategy after such reports, since they do not figure in the quack's indifference condition. Indeed, we show that the judge's selection after two inconsistent messages also does not affect the optimality of truth-telling for the expert: reporting the true state is optimal for the expert, even if the judge were to miraculously select him for sure after two inconsistent reports.

\subsubsection{Best Equilibrium for the Judge}\label{sec:arg3}

Combining our results in Sections \ref{sec:arg1} and \ref{sec:arg2}, we know that an equilibrium exists in which the expert uses messaging strategy $F_E(\cdot|\omega)=\delta_\omega$, for each $\omega$ in $[-1,1]$, the quack uses strategy $F_Q=\mathcal{U}[-1,1]$, and the judge uses a selection rule $\phi$ in $\Phi^{ind}$. We further show that this equilibrium maximizes the judge's payoff.

\begin{proposition}\label{pr:1}
The equilibrium payoff of the judge is maximal in any equilibrium
in which the expert reports the state truthfully.
\end{proposition}

The proof of Proposition \ref{pr:1}  is a specialization of Proposition \ref{pr:generalbest}, which shows that, in a more general environment, the truth-telling equilibrium attains the payoff bound of the value the judge can attain when committing to a mechanism to incentivize reports from the two speakers, about the state and about their expertise types. This should come as no surprise, since commitment has no value in a zero-sum game, which is what the game reduces to, once the expert's strategy is fixed. 

A further implication of our argument is that, because the truth-telling equilibrium in the simultaneous-speech game attains the commitment-payoff bound, it also delivers the best payoff to the judge across equilibria of all possible extensive-form games in this environment: more sophisticated ``debates'' between the two speakers cannot help the judge.

\subsection{Learning the State}\label{sec:features}
In the equilibrium described in Theorem \ref{thm:best}, the expert reports the state truthfully, but the judge need not recognize which message is correct.  If only one message is consistent with her signal, the judge identifies the expert and learns the state with certainty. If both messages are consistent, her posterior assigns equal probability to the two reported states. Communication therefore narrows the judge's uncertainty to at most two possibilities, even when it does not resolve that uncertainty completely.

In the context of the truth-telling equilibrium, we examine how the precision of the judge's private signal affects their learning about the state. We consider two measures of learning: the probability of learning the state with certainty and the reduction in expected posterior variance due to communication. These measures capture different benefits of hearing the speakers and yield different conclusions about which judges benefit most from communication. We then compare learning across equilibria, using the probability that a judge's guess of the state is exactly correct.

\paragraph{Signal Accuracy and Learning the State.} The precision of the judge's private signal determines the likelihood that she learns the state perfectly: for a given draw of the private signal $s$, a given message $m$ sent by a speaker is considered consistent if $|m-s|\leqslant\bar{\epsilon}$, so that the higher the noise component in the judge's signal, the less restrictive the condition that determines a message's consistency. Proposition \ref{pr:2} describes how often and when the judge learns the state perfectly, relating this to the precision of her private signal. To that end, let $\hat{G}$ in $\Delta[-1,1]$ be the judge's posterior belief about the state.

\begin{proposition}\label{pr:2}
The probability that the judge learns the state is
$\mathbf{P}\left(\hat{G}=\delta_\omega\right)=1-\bar{\epsilon}+\frac{\bar{\epsilon}^2}{3}$,
which is increasing in the precision of the judge's private signal.
 Conditional on a particular realization of the state, the judge learns the state with probability
$$\mathbf{P}\left(\hat{G}=\delta_\omega\mid\omega=\hat{\omega}\right)=1-\bar{\epsilon}
+\frac{(\hat{\omega}+2\bar{\epsilon}-1)_+^2+(-\hat{\omega}+2\bar{\epsilon}-1)_+^2}{8\bar{\epsilon}},$$
where $x_+:=\max\{x,0\}$. Therefore, the judge is weakly more likely to learn more extreme states.

Moreover, for every $\bar{\omega}$ in $(0,1)$, the conditional probability
\[
\mathbf{P}\left(
|\omega|>\bar{\omega}
\mid
\hat{G}=\delta_\omega
\right)
\]
is increasing in $\bar\epsilon$ on $(0,1]$. Consequently, although less informed judges learn the state with certainty less often, conditional on doing so they are more likely to learn extreme states.
\end{proposition}

Proposition \ref{pr:2} measures learning by the probability that the judge becomes certain of the state. Communication also provides information when both messages remain consistent with her private signal: the judge then knows that the state is one of the two reported values. To account for this partial learning, we consider the reduction in expected posterior variance due to communication, relative to observing the private signal alone. Proposition \ref{pr:var} shows that this reduction is larger for judges with less precise private signals, yielding a different comparison of the informational benefits of communication.

\begin{proposition}\label{pr:var}
The reduction in expected posterior variance is given by
$$\mathbf{E}\left[\mathrm{Var}(\omega|s)-\mathbf{E}\left(\mathrm{Var}(\omega|s,m_1,m_2)|s\right)\right]=\frac{1}{3}\bar{\epsilon}^2-\frac{1}{3}\bar{\epsilon}^3+\frac{1}{10}\bar{\epsilon}^4,$$
which is increasing in $\bar{\epsilon}$ on $[0,1]$ (decreasing in the precision of the judge's private signal).
\end{proposition}

\paragraph{Learning and Informational Inequality.} Proposition \ref{pr:2} highlights a sense in which information begets information. A more precise private signal enables the judge to rule out the quack’s report more often and thereby more often recognize the truthful report and learn the state.
These differences in learning across judges with different signal precisions arise not from differences in speakers' communication strategies in equilibrium, but from  the judges' distinct ability to evaluate speakers' reports.
Through this mechanism, we find that exposure to communication from speakers of unknown expertise can amplify informational inequality in a population of judges with different signal precision levels:\footnote{Our equilibrium construction extends to the case in which there is a judge with unknown noise, or alternatively a population of judges with conditionally independent private signals, perhaps with different noise levels (in this case, the speakers' objectives are to maximize the number of judges that select them).} Before communication, no judge learns the state exactly from her private signal alone, so that there is no difference in their ability to exactly guess the state. In turn, after communication, better informed judges are more likely to know the state with certainty.

Whether communication widens or narrows informational differences, however, depends on the measure we use to assess the value of information. Proposition \ref{pr:var} shows that less informed judges obtain larger reductions in expected posterior variance from communication. Although they are less likely to resolve their uncertainty completely, the information conveyed by the speakers produces a larger absolute reduction in their expected uncertainty. Consequently, communication narrows the difference in expected posterior variance between judges with different signal precisions. The advantage of prior information in attaining certainty thus coexists with a narrowing of informational differences measured by expected posterior variance.

Signal accuracy also affects the composition of what judges learn. Proposition 2 shows that, for any given precision, extreme states are weakly more likely to be learned than moderate states. Moreover, conditional on learning the state with certainty, judges with less precise signals learn a distribution of states that is more concentrated toward the extremes. Less informed judges therefore learn less often, and the occasions on which they do learn disproportionately concern extreme realizations.

The composition of successful learning differs from selection conditional
on a quack's report being consistent. Equation \eqref{eq:p2} shows that
more extreme reports have a weakly higher conditional selection
probability. Comparisons of this probability across noise levels depend
on the report.

\paragraph{Learning the State Across Equilibria.} Theorem \ref{thm:best} shows that the expert-truth-telling equilibrium maximizes the probability that the judge selects the expert. Proposition \ref{pr:8} connects the judge's equilibrium optimality  to the distinct objective of accurately estimating the state. It shows that the probability that the judge accurately estimates the state $\omega$ is bounded above in equilibrium by the probability that the judge selects the expert. Moreover, this bound is attained at the truth-telling equilibrium, so that the maximization of the judge's expert-identification objective is aligned with that of estimating the state.\footnote{Proposition \ref{pr:8} bounds the probability of a correct estimate across equilibria. This does not establish a Blackwell ordering of the information learned by the judge. Online Appendix OA.7 presents numerical evidence against such an ordering in a finite version of the game.}

Formally, an \emph{estimate} (of the state) is a Borel map $\hat{g}:[-1,1]^2\times\mathbf{R}\rightarrow\mathbf{R}$, associating with the judge's information --- the pair of reports $(m_1,m_2)$ and her signal $s$ --- a guess $\hat{g}(m_1,m_2,s)$ of $\omega$. An estimate is an auxiliary object, not a move of the game: it affects no payoff, and the judge is free to form one alongside her selection. Its accuracy is evaluated under the joint distribution induced by the equilibrium.

\begin{proposition}\label{pr:8}
In any equilibrium of the communication game, and for every estimate $\hat{g}$,
\[
\mathbf{P}\left[\hat{g}(m_1,m_2,s)=\omega\right]\leqslant 1-\frac{\bar\epsilon}{2}+\frac{\bar\epsilon^2}{6}.
\]
Moreover, every equilibrium in which the expert reports the state truthfully admits an estimate that attains this bound.
\end{proposition}

The proof, presented in Appendix \ref{sec:proppr8}, rests on the following logic:
Any guess of the state that is correct with positive probability must decode the message of one of the two speakers: the judge's signal $s$, which admits a density, cannot by itself single out $\omega$, and the quack's message, being independent of $\omega$, almost surely never reveals its exact value. A correct guess therefore permits a selection rule that selects the expert on that event. Consequently, the probability of correctly estimating the state is bounded by the judge's probability of selecting the expert, which is itself maximal in truth-telling equilibria.

\subsection{Nonuniform Prior}\label{sec:state}
In our baseline specification, the state is distributed according to the uniform distribution over the interval $[-1,1]$. We now consider a variation in which the state distribution is thinner-tailed. As in our baseline, an equilibrium exists in which the expert truthfully reports the state, and this equilibrium attains the judge's maximal payoff.

However, the modification affects the quack's incentives to mimic the expert. The quack's equilibrium strategy is to now \emph{pander to moderate messages}  --- those that are more likely to be consistent with the judge’s signal about the state. Proposition \ref{pr:3} specifies the shape of the quack's equilibrium pandering.

\begin{proposition}\label{pr:3}
Suppose $\omega\sim G$, where $G$ has a log-concave density on its support $[-1,1]$ and is symmetric around $0$. There exists an equilibrium in which the expert truthfully reports the state, and the quack mimics the prior over moderate messages. That is, there exists some threshold $\bar{m}$ in $(0,1]$ such that the quack's messaging strategy is
$$F_Q(m)=\begin{cases}
0\text{, if }m\leqslant -\bar{m},\\
\frac{G(m)-G(-\bar{m})}{G(\bar{m})-G(-\bar{m})}\text{, if }m\in[-\bar{m},\bar{m}],\\
1\text{, if }m\geqslant\bar{m}.\end{cases}$$
\end{proposition}

When $G$ has sufficiently thin tails, the quack uses a messaging strategy with a truncated support, that avoids extreme messages and overweights moderate messages. Conditional on a message belonging to a sufficiently moderate interval, the quack's strategy exactly mirrors the prior state distribution, as well as the expert's marginal messaging strategy.

When $\bar m<1$, the boundary report yields the quack's equilibrium payoff if it is selected whenever it is consistent and the expert's report lies in $[-\bar m,\bar m]$. An expert report outside this interval identifies the expert. The first-order condition in the proof of Proposition \ref{pr:3} equates this boundary payoff to the payoff from reports inside the support. When $\bar m=1$, the optimum is a boundary solution and this equality need not hold.

\subsection{Unequal Identity Priors}\label{sec:identity}
We now consider a variation of the model in which the judge has some information about the identity of the expert. Suppose the judge believes speaker $i$ to be the expert with probability $p_i$ in $(0,1)$, for $i$ in $\{1,2\}$; and let $p_1>1/2>p_2$. We accordingly refer to speaker $1$ as the favored speaker, and speaker $2$ as the disfavored speaker.

Our equilibrium construction now involves not only a messaging strategy for the expert and one for the quack, but instead potentially distinct quack and expert messaging strategies for each of the two speakers. We let $F^i_E$ in $\Delta([-1,1]\times\mathcal{M})$ and $F^i_Q$ in $\Delta\mathcal{M}$ be the messaging strategies used by speaker $i$ in $\{1,2\}$ when he is the expert or the quack, respectively. Proposition \ref{pr:4} describes an equilibrium in which each speaker truthfully reports the state when they are the expert.\footnote{As in our benchmark, this equilibrium attains the judge's commitment-payoff upper bound. The proof of Proposition \ref{pr:4} also describes the judge's expected error probability, which is lower when the judge has more information about the speaker's identities (that is, if $p_1$ is larger). } In contrast, conditional on being the quack, the favored and disfavored speakers behave differently, with the favored speaker using a conservative messaging strategy that panders to moderate messages.

Set $\bar m=\max\{p_2/p_1,(1-2\bep)_+\}$.

\begin{proposition}\label{pr:4}
Suppose $p_1>1/2>p_2$. There is an equilibrium in which each expert reports
the state truthfully. The disfavored quack uses $\mathcal U[-1,1]$.
The favored quack uses $\mathcal U[-\bar m,\bar m]$.
If $p_2/p_1<(1-2\bep)_+$, any
$\bar m$ in $[p_2/p_1,1-2\bep]$ can instead be used.
\end{proposition}

Call the asymmetry \emph{extreme} when $p_2/p_1\le(1-2\bep)_+$, and \emph{moderate} otherwise. If the asymmetry is extreme, the favored speaker is selected by the judge every time his message is consistent, regardless of the message used by the other speaker. Given this selection rule, the favored quack's messaging strategy must use only ``moderate enough'' messages, which maximize the probability of consistency. Similarly, when the asymmetry is less extreme, the judge is more willing to select the favored speaker, which yields an equilibrium  in which the favored quack panders (partially) to messages that are more moderate. In this equilibrium, the disfavored quack does not pander.

\section{Applications}\label{sec:applications}
\subsection{AI Evaluation}

AI chatbots are shaped both by pre-training, which produces a base model from large amounts of data, and by subsequent post-training, which steers the model toward desired patterns of behavior. Our application speaks primarily to this latter stage. A prominent class of post-training procedures, including reinforcement learning from human feedback (RLHF), relies on comparative feedback: human evaluators rank or choose between responses to the same query, and these comparisons are then used to construct reward signals that favor responses evaluators regard as better. Our model studies the incentives generated by such a comparative reward structure when evaluators seek to identify the response that is better informed.

Specifically, we interpret the expert speaker as a grounded system, one that has access to the retrieval source, data, or tool needed for the query at hand. The quack, in contrast, is an ungrounded system, which lacks this information but can nevertheless strategically produce a plausible response. The judge is a human evaluator who wishes to identify the response produced by the better-informed system. Her selection may determine the reward used to post-train the model or which system is ultimately deployed. The query therefore serves as an evaluation task through which the human tries to infer which system has access to the relevant information. Although she cannot observe this directly, she has some imperfect knowledge about the answer herself. We interpret her private signal as background knowledge, domain-specific heuristics, or partial information that allows her to assess whether a candidate response is plausible without knowing the correct answer with certainty.

The speakers’ payoffs capture the incentives induced by comparative evaluation: systems are rewarded for producing responses that evaluators select. They are not adversarial to the evaluator, nor do they seek to mislead her; any strategic behavior arises solely from this reward for being selected. We view post-training as the leading interpretation, though the same structure can also describe the evaluation of competing systems for deployment.

\paragraph{Anti-Sycophancy and Truthful Reporting.} The first implication of Theorem \ref{thm:best} is that comparative evaluation can sustain truthful reporting by the grounded system. Moreover, our construction shows the importance of feedback from a human evaluator who actively chooses which system to reward. Suppose instead that rewards were mechanically tied to the evaluator's posterior belief that each system is informed. Then, whenever the two responses are plausible (consistent), they have to receive the same reward, as in our ``uniform tie-breaking'' candidate rule. As we have argued, this creates an incentive for the ungrounded system to pander toward moderate answers, which are \textit{ex ante} more likely to be consistent with the evaluator's private information.

In the AI chatbot context, we interpret this incentive as a form of \emph{sycophancy}: a system can improve its chances of being rewarded by producing answers that the evaluator is already predisposed to accept. Our equilibrium counteracts this incentive through the evaluator's selection behavior. When both answers are consistent with her private information and she is therefore indifferent between the systems, she favors the more ``extreme,'' and hence less sycophantic, response. This conditional premium on less sycophantic answers offsets the greater \textit{ex ante} plausibility of moderate responses. As a result, the ungrounded system randomizes across answers rather than systematically pandering, and truthful reporting is optimal for the grounded system.

The stark distinction between a perfectly informed grounded system and an uninformed system is not essential to this conclusion. Section \ref{sec:generalmodel} shows that truthful reporting can also be sustained when both systems are imperfectly informed, provided the less informed system observes a regular garbling of the information available to the more informed one.

\paragraph{The Source and Shape of Hallucinations.} Because the grounded system produces only truthful answers in our equilibrium, our model advances the theory that hallucinations originate with the ungrounded system, as a strategic attempt to be selected by the evaluator. These hallucinations by the ungrounded system cannot always be detected by the user, but our theory predicts some features of the distribution of hallucinations, as well as how they change as we depart from our benchmark.

In the benchmark, the unconditional distribution of hallucinations exactly replicates the marginal distribution of truthful answers; hallucinations do not pander towards or against the user's prior, in the aggregate. Our extensions yield two systematic departures from this benchmark, which generate empirical predictions relating the distribution of hallucinations --- and in particular their degree of sycophancy --- to features of the evaluation environment.

Specifically, hallucinations become more concentrated on safe, sycophantic answers when (i) correct answers are themselves tightly concentrated, as in the thin-tailed state distribution of Section \ref{sec:state}, and (ii) one of the systems begins with a stronger reputation for expertise, as in Section \ref{sec:identity}. In particular, an initially favored but ungrounded system should hallucinate more conservatively than an initially disfavored one.\footnote{In principle, these predictions can be tested by varying the query domain or the disclosed identity and reputation of the system, and examining whether its errors shift toward answers users were already predisposed to accept.} In Section \ref{sec:generalmodel} and Online Appendix OA.4, we discuss a third departure from the benchmark, establishing that under smooth noise all reports remain in use while hallucinations shift toward more moderate, sycophantic answers.\footnote{The form of  pandering differs across these environments. In the extensions considered in Sections \ref{sec:state} and \ref{sec:identity}, pandering operates on the extensive margin: sufficiently extreme reports are dropped from the ungrounded system's support. With smooth noise in the evaluator's private signal, Online Appendix OA.4 establishes that all reports remain in use. Their relative frequencies tilt toward more moderate answers.}

\paragraph{Optimal Comparative Evaluation.} Within our environment, the truthful equilibrium of comparative evaluation attains the highest feasible probability of selecting the grounded system. This screening objective is inseparable from learning about the query, and the same equilibrium also maximizes the probability that the user's estimate of the answer is correct.

Moreover, the user cannot improve on this outcome by committing \textit{ex ante} to a different reward or selection rule. The absence of commitment gains has a dynamic-consistency interpretation: the optimum can be implemented by having the user select the system that is most likely to be grounded \emph{after} observing the responses and her private signal. The anti-sycophantic selection bias required to sustain truth-telling operates only when the user is indifferent between the two answers. The protocol therefore never requires the user to reward an answer that she regards as less credible.

\paragraph{User Knowledge and Learning from AI.} The truthful equilibrium can be sustained for any precision of the user's background knowledge.  User expertise is a determinant of equilibrium performance, rather than a prerequisite for truthful communication: better-informed users are more likely to distinguish grounded answers from hallucinations and thereby learn the correct answer. Access to the same AI systems can consequently complement the information users already possess and amplify differences in users' ability to identify the correct answer.

Signal accuracy also changes the selection rule. Under the max rule,
the difference between the selection probabilities assigned to the most
and least extreme reports increases with accuracy. The rules need not
be ordered at intermediate reports.

\subsection{Media Competition}

Media competition provides a second application of our model. Here, the speakers are media sources or commentators presenting conflicting accounts of the same issue, and the judge is an audience member deciding which source to trust. One source has access to the information needed to report accurately, while the other is uninformed. The audience member draws on imperfect background knowledge to decide which account to believe and which source to follow or consult in the future. Each source therefore competes for the audience's attention and for their readership or subscriptions which come from being regarded as reliable.

\paragraph{Source Reliability and Information Inequality.} In a media environment with plentiful sources of uncertain reliability, an audience member's background knowledge --- reflecting their education or experience --- affects how much they can learn from new reports. A better informed audience member can use that knowledge to distinguish informed reporting from plausible imitation and thereby recover the information contained in the informed report more often. Exposure to the same sources therefore need not level initial informational differences across audiences, but can instead amplify them. This mechanism offers an explanation for the coexistence of widespread access to information and persistent or growing information inequality.

Prominent theories of information inequality instead emphasize ``partisan echo chambers.'' In Gentzkow and Shapiro (2006), media companies pander to their audiences' priors. Perego and Yuksel (2022) identify a different channel: competition induces outlets to specialize in issues on which audiences' preferences differ. An implication is that information polarization or inequality can arise because different media outlets cater to pre-existing heterogeneous priors across a population. In our model, all audiences are exposed to the same sources, which do not pander in our benchmark equilibrium; informational inequality arises because audiences differ in their ability to assess the reliability of the same sources.

Evidence from Angelucci and Prat (2024) is consistent with this channel.
In their survey, subjects are confronted with true and fake news stories regarding the same major political events. They show that there is significant variability in subjects' abilities to pick the true story from the false, and investigate potential sources of this variability. While partisan congruence between an individual and a news story matters for their choice of truth (as in the common ``echo chamber story''), they show that the impact of partisanship is up to an order of magnitude smaller than that of differences in participants' socioeconomic backgrounds, which plausibly proxy for the prior information or domain knowledge differences that support our proposed mechanism.

The sense in which communication amplifies informational inequality nevertheless depends on how learning is measured. Better-informed audiences are more likely to learn the state with certainty, but less-informed audiences experience a larger reduction in expected posterior variance from communication. Thus, while communication widens differences in audiences' ability to identify the truth exactly (Proposition \ref{pr:2}), it can narrow differences in their residual uncertainty (Proposition \ref{pr:var}).

\paragraph{Audience Knowledge and the Composition of Learning.} Proposition \ref{pr:2} predicts that the audience's background knowledge affects not only how often they learn, but the composition of what they learn. Conditional on learning the state with certainty, less-informed audiences are disproportionately likely to learn more ``extreme'' states --- that is, realizations that are \textit{ex ante} less likely to be viewed as plausible. In the media context, this implies that, although less-informed audiences correctly identify the truth less often overall, their successful learning should be relatively concentrated on unusual news that is \textit{ex ante} less likely to be viewed as plausible. This provides a potentially testable distinction from theories in which informational differences primarily arise from partisan sorting or pandering to heterogeneous priors: holding exposure to news sources fixed, our mechanism predicts systematic differences in the composition of correctly learned news according to audiences' prior knowledge, rather than according to partisan predispositions.

\section{Ordered Expertise}\label{sec:generalmodel}

Our benchmark environment and its extensions maintain two strong assumptions for tractability. First, the judge's signal about the state has a uniformly distributed error, which implies that the judge can use her private information only to separate ``possibly true'' states from those that are inconsistent. Second, we take a stark view of expertise:  the expert observes the state perfectly, and the quack is perfectly uninformed.

These assumptions are not needed for truthful communication. In this section, we treat expertise as an ordering of information. The expert observes a signal, and the quack observes a garbling of that signal. The words ``expert'' and ``quack''  continue to distinguish the speakers' roles, but they mean more versus less informed, respectively.

\paragraph{Environment.} Formally, the state $\omega$ has an arbitrary Borel distribution $G$ on $[-1,1]$. The expert observes a signal of the state,
\[
\theta=\omega+\epsilon_E,
\]
and the quack observes a signal $t$, which is a garbling of that of the expert, drawn according to a Borel kernel $K(\de t\mid\theta)$ on $\mathbf R$. Finally, the judge's private signal is
\[
s=\omega+\epsilon_J.
\]
The noises $\epsilon_E$ and $\epsilon_J$ are independent of each other and of the state, with Borel densities $h_E$ and $h_J$ on $\mathbf{R}$, respectively. Conditional on $\theta$, the garbling that yields the quack's signal is independent of the state and of the judge's signal. As before, there is exactly one expert and one quack; the role assignments are known by the speakers, whereas the judge deems each speaker equally likely to be the expert or the quack.

The order of play, as well as the payoff structure, is unchanged from the benchmark. Each speaker's strategy now maps their private signal to a distribution of messages over $\mathbf{R}$.

When we refer to ``truth-telling,'' we mean the strategy in which the expert speaker reports their true signal $\theta$ --- as opposed to reporting the true state $\omega$, in the benchmark. Accordingly, the entire model is most transparently expressed around the expert's signal. Its law is
\[
\Gamma(\de\theta)=\gamma(\theta)\de\theta,
\qquad\text{with }
\gamma(\theta):=\int h_E(\theta-\omega)G(\de\omega).
\]
Conditional on $\theta$, the judge's signal has density
\[
\hat{h}(s\mid\theta):=\int h_J(s-\omega)G_\theta(\de\omega),
\qquad\text{where }
G_\theta(\de\omega):=\frac{h_E(\theta-\omega)}{\gamma(\theta)}G(\de\omega)
\]
whenever $\gamma(\theta)>0$.\footnote{More generally, the following reduced model may be taken as primitive: $\Gamma$ is any Borel law of the expert's signal, and $\hat{h}(\cdot\mid\theta)$ any jointly Borel family of probability densities for the judge's signal. This formulation includes perfect expert information, for which $\theta=\omega$, $\Gamma=G$, and $\hat{h}(s\mid\theta)=h_J(s-\theta)$, even though the corresponding error $\epsilon_E\equiv0$ does not have a density.}
The joint law of the expert's type and the quack's signal is $\Gamma(\de\theta)K(\de t\mid\theta)$. Alternatively, we write $\lambda$ for its $t$-marginal and $\Gamma_t$ for a regular conditional distribution of $\theta$ given $t$, so that
\[
\Gamma(\de\theta)K(\de t\mid\theta)=\Gamma_t(\de\theta)\lambda(\de t).
\]

Note that, conditional on the expert's signal $\theta$, the quack's signal is independent of the state and of the judge's signal. Thus the information structure forms the Markov chain $\omega\rightarrow\theta\rightarrow t$; the judge's signal is conditionally independent of the quack's signal given $\theta$.

We assume the following \emph{regularity} condition on the garbling $K$: there is a subset $O$ of $\mathbf R$ of full $\lambda$-measure such that $\lambda(V)>0$ for every nonempty relatively open subset $V$ of $O$, and the posterior map $t\mapsto\Gamma_t$ admits a version that is continuous in total variation on $O$. This is a condition on the reduced information structure (not a shape restriction), which states that nearby quack signals induce nearby beliefs about the expert's signal.\footnote{For an uninformed quack, $\Gamma_t\equiv\Gamma$, so regularity is immediate without any restriction on $\Gamma$. The benchmark model is the reduced specification with $\Gamma=\mathcal U[-1,1]$, $\hat{h}(s\mid\theta)=(2\bar\epsilon)^{-1}\mathbf{1}\{|s-\theta|\leqslant\bar\epsilon\}$, and $K(\cdot\mid\theta)=\delta_0$. }

\paragraph{Truth-Telling Game.}
We refer to the zero-sum two-player game between the judge and the quack implied when the expert's strategy is  truth-telling as the \emph{truth-telling game}.

A quack strategy for speaker $i$ in $\{1,2\}$ is a Borel kernel $F_i(\cdot\mid t)$ from signals to reports. The judge uses a Borel selection rule $\phi$ with values in $[0,1]$, which indicates the probability of selecting speaker 1. If speaker $i$ is the quack and observes $t$, his probability of being selected by the judge when using message $m$ is
\begin{equation}\label{eq:generalQ}
Q_i^{\phi}(m\mid t)
:=\int\!\int q_i(m,m',s)\hat{h}(s\mid m')\de s\,\Gamma_t(\de m'),
\end{equation}
where $q_1(m,m',s)=\phi(m,m',s)$ and $q_2(m,m',s)=1-\phi(m',m,s)$.

The quack's objective is to maximize --- and the judge's is to minimize --- the \textit{ex ante} probability that the quack is selected. Their interaction is therefore zero-sum. The quack's \textit{ex ante} expected payoff (averaging also over the identity assignment) is
\[
U(F_1,F_2,\phi):=\frac12\sum_{i=1}^2
 \int_{\mathbf R}\!\int_{\mathbf R}Q_i^{\phi}(m\mid t)
 F_i(\de m\mid t)\lambda(\de t).
\]

\subsection{Truthful Communication}\label{sec:generalresults}

Theorem \ref{thm:generaltruth} establishes the existence of a saddle point in the truth-telling game between the quack and the judge. Moreover, for any such saddle point, we verify that the expert does not have incentives to deviate from the truth-telling strategy that was fixed, so that the saddle point extends to a truth-telling equilibrium of the original three-player game.\footnote{\label{foot:1}The existence and truth-telling conclusions continue to hold with arbitrary strictly positive prior probabilities on the two identity assignments. If those probabilities are equal and the reduced primitives are symmetric --- $\Gamma$ is symmetric, $\hat{h}(-s\mid-\theta)=\hat{h}(s\mid\theta)$, and $K(-A\mid-\theta)=K(A\mid\theta)$ for every Borel $A$ --- a symmetric saddle point exists. The proof of this statement is included in the proof of the theorem.}

\begin{theorem}\label{thm:generaltruth}
Assume that the garbling kernel $K$ is regular. The truth-telling game has a (quack) value $v$ and a saddle point. At every saddle point of the truth-telling game, truthful reporting is a best response for the expert for $\Gamma$-almost every signal $\theta$ in the original three-player game. Every saddle point therefore extends, after completing off-path beliefs, to a truth-telling equilibrium of the three-player game.
\end{theorem}

The proof of Theorem \ref{thm:generaltruth} first shows that a saddle point of the truth-telling game exists; this argument uses regularity, which protects the judge against deviations that correlate exceptional reports with the quack's signal. Conditional on existence, the expert's truth-telling incentives rest only on the conditional-independence structure and the common dominating measure for the judge's conditional signal laws.

The argument that extends the saddle point into a truth-telling equilibrium of the three-player game is an aggregated version of the one in our benchmark: At a saddle point, the quack's reports equalize his payoff conditional on each signal $t$. The expert's gain from replacing the truthful report $\theta$ by some $m$ is bounded by the quack's expected gain from the same replacement, averaged over precisely the quack signals at which he sends $\theta$. That average is nonpositive by conditional equalization. The quack's information changes the composition of the average, but not its sign. Truth-telling is therefore guaranteed not by the quack's ignorance, but by the fact that the expert directly observes the variable the quack is trying to infer.

This nested structure is substantive. If the expert and the quack instead observe independent noisy signals of the state, then, conditional on $\theta$, the quack's signal remains correlated with the state and hence with the judge's signal. The aggregated comparison no longer applies.

Proposition \ref{pr:generalbest} generalizes our benchmark result --- in Proposition \ref{pr:1} --- that the truth-telling equilibrium attains the judge's commitment-payoff bound.

\begin{proposition}\label{pr:generalbest}
Every truth-telling saddle point gives the judge payoff $1-v$, which is the maximal payoff attainable by the judge among equilibria of all communication mechanisms.
\end{proposition}

\subsection{Pandering}
Theorem \ref{thm:generaltruth} establishes truthful communication under a general ordering of expertise, generalizing the observation from our benchmark model, but places little structure on the shape of the quack’s equilibrium reports. Online Appendix OA.4 studies the quack's messaging strategies in the expert-truth-telling equilibria, in settings beyond our benchmark. Under the smooth-noise and prior assumptions stated there, these strategies overweight moderate messages, thereby ``pandering'' to states the judge is likely to deem more plausible.

This pandering feature is similarly discussed in  Sections \ref{sec:state} and \ref{sec:identity}, where we discuss extensions of the benchmark model. Specifically, in those extensions, pandering operates on the \emph{extensive} margin: the quack drops sufficiently extreme reports and reproduces the expert’s marginal report distribution over a central interval. Under smooth noise of the judge's signal, all reports remain in use. Their relative frequencies shift toward the center, so pandering operates on the \emph{intensive} margin.

Concretely, in Online Appendix OA.4,  we consider a setting in which the quack is fully uninformed, the expert is perfectly informed, and the judge's signal has noise satisfying Assumption 1 of that appendix.\footnote{The noise has a smooth, symmetric distribution with thin tails, so extreme errors are rare but every signal remains possible.} The state prior satisfies Assumption 2: its density is smooth, positive, symmetric, and nonincreasing away from the center. Proposition OA.4.3 proves that the quack's equilibrium report density $f_Q$ is positive throughout the state interval. The ratio $f_Q/g$ decreases strictly with the absolute value of the report. Thus the quack uses every report but assigns more probability than the truthful expert to every central interval. This pandering is bounded: $f_Q/(gh)$ increases strictly with the absolute value of the report, so the quack's messages are less concentrated toward the center than the posterior of a judge whose signal is central. Corollary OA.4.4 shows how the judge offsets these incentives. Her selection cutoff favors the more extreme report, but she still selects the more moderate report when her signal is central.

\section{Concluding Remarks}\label{sec:conclusion}

Throughout the paper, we kept the structure of the communication game unchanged: there are two speakers, whose expertise types are anti-correlated, and speakers report to the judge simultaneously. One of our results, Proposition \ref{pr:1}, argued that, even if the judge could alter the extensive form of the communication game, she would not be able to better screen for the quack, compared to simultaneous communication. In Online Appendix OA.5, we illustrate one possible alternative extensive form, in which speakers report sequentially, and openly (so the second speaker can see the report of the first speaker before reporting himself). For $\bep<1/4$, every truth-telling equilibrium of that sequential game gives the judge a lower payoff than the truth-telling equilibrium with simultaneous reports.

Online Appendix OA.6 compares the reputational cheap-talk game played by two speakers with anti-correlated types to the alternative setting with a single speaker, where the judge selects between the speaker and some fixed outside option. As in our main model, we construct equilibria in which the speaker reports the true state, when he is the expert. In contrast with the two-speaker scenario, with one speaker, the judge evaluates absolute plausibility rather than relative plausibility. Because more extreme messages are stronger conditional signals of expertise, the quack sometimes overweights extremes, a behavior absent from our benchmark and excluded under the smooth-noise assumptions in Online Appendix OA.4.


Our analysis leaves other alternative specifications and questions for future research: What if there are more than two speakers? Does competition among quacks lead to more extremism in their stances? Is competition beneficial for learning by the judge? Also, speakers are rarely evaluated according to a single dimension. Politicians, media outlets and LLMs might be more or less competent depending on the topic at hand. There might be benefits from breaking the debate down into different themes before making a choice. Finally, speakers' types might not be correlated,  and if they are, it might be beneficial to have a structured debate where speakers might get the opportunity to ask questions to one another, calling them out, as their superior information might precisely be about  where the ignorance of others lies.
\section*{References}

\textbf{Alysandratos, T., A. Boukouras, S. Georganas, Z. Maniadis, and A. Velias} (2023).
``Science versus Populism: Why and When Do Charlatans Beat Experts?''
Working paper, October 6.

\textbf{Andreoni, J., and T.\ Mylovanov} (2012).
``Diverging Opinions,''
\textit{American Economic Journal: Microeconomics}, \textbf{4}, 209--232.

\textbf{Angelucci, C., and A.\ Prat} (2024).
``Is Journalistic Truth Dead? Measuring How Informed Voters are about Political News,''
\textit{American Economic Review}, \textbf{114}, 887--925.

\textbf{Battaglini, M.} (2002).
``Multiple Referrals and Multidimensional Cheap Talk,''
\textit{Econometrica}, \textbf{70}, 1379--1401.

\textbf{Bertsekas, D. P., and S. E. Shreve} (1978).
\textit{Stochastic Optimal Control: The Discrete-Time Case}.
Academic Press.

\textbf{Brandenburger, A., and B.\ Polak} (1996).
``When Managers Cover Their Posteriors: Making the Decisions the Market Wants to See,''
\textit{The RAND Journal of Economics}, \textbf{27}, 523--541.

\textbf{Chen, E. O., A. Ghersengorin, and S. Petersen} (2024). ``Imperfect Recall and AI Delegation,'' Global Priorities Institute Working Paper 30-2024.

\textbf{Crawford, V., and J.\ Sobel} (1982).
``Strategic Information Transmission,''
\textit{Econometrica}, \textbf{50}, 1431--1451.

\textbf{Dworczak, P. and A. Smolin} (2026). ``Robust Trust,'' \textit{working paper}.

\textbf{Friedman, E.} (1998).
``Public Debate among Experts,''
Northwestern University, CMS-EMS Working Paper 1234.

\textbf{Fudenberg, D. and A. Liang} (2026). ``Friend or Foe: Delegating to an AI whose Alignment is Unknown,'' \textit{working paper}.

\textbf{Gentzkow, M., and J.\ M.\ Shapiro} (2006).
``Media Bias and Reputation,''
\textit{Journal of Political Economy}, \textbf{114}, 280--316.

\textbf{Glazer, J., and C.-T.\ Ma} (1989).
``Efficient Allocation of a `Prize'---King Solomon's Dilemma,''
\textit{Games and Economic Behavior}, \textbf{1}, 222--233.

\textbf{Glazer, J., and A.\ Rubinstein} (1998).
``Motives and Implementation: On the Design of Mechanisms to Elicit Opinions,''
\textit{Journal of Economic Theory}, \textbf{79}, 157--173.

\textbf{Glazer, J., and A.\ Rubinstein} (2001).
``Debates and Decisions: On a Rationale of Argumentation Rules,''
\textit{Games and Economic Behavior}, \textbf{36}, 158--173.

\textbf{Green, J.\ R., and N.\ Stokey} (2007).
``A Two-Person Game of Information Transmission,''
\textit{Journal of Economic Theory}, \textbf{135}, 90--104.

\textbf{Huang, L., W. Yu, W. Ma, W. Zhong, Z. Feng, H. Wang, Q. Chen \textit{et al.}} (2025). ``A survey on hallucination in large language models: Principles, taxonomy, challenges, and open questions.'' \textit{ACM Transactions on Information Systems}, \textbf{43}(2), Article 42, 1--55.

\textbf{Krishna, V., and J.\ Morgan} (2001).
``A Model of Expertise,''
\textit{The Quarterly Journal of Economics}, \textbf{116}, 747--775.

\textbf{Lehmann, E. L.} (1988).
``Comparing Location Experiments,''
\textit{Annals of Statistics}, \textbf{16}, 521--533.

\textbf{Levy, G.} (2004).
    ``Anti-herding and Strategic Consultation,''
    \textit{European Economic Review}, \textbf{48}, 503--525.

\textbf{Maskin, E.} (1999).
``Nash Equilibrium and Welfare Optimality,''
\textit{Review of Economic Studies}, \textbf{66}, 23--38.

\textbf{Morris, S.} (2001).
``Political Correctness,''
\textit{Journal of Political Economy}, \textbf{109}, 231--265.

\textbf{Olszewski, W.} (2004).
``Informal Communication,''
\textit{Journal of Economic Theory}, \textbf{117}, 180--200.

\textbf{Ottaviani, M., and P.\ S\o rensen} (2001).
``Information Aggregation in Debate: Who Should Speak First?,''
\textit{Journal of Public Economics}, \textbf{81}, 393--421.

\textbf{Ottaviani, M., and P.\ S\o rensen} (2006a).
``The Strategy of Professional Forecasting,''
\textit{Journal of Financial Economics}, \textbf{81}, 441--466.

\textbf{Ottaviani, M., and P.\ S\o rensen} (2006b).
``Reputational Cheap Talk,''
\textit{The RAND Journal of Economics}, \textbf{37}, 155--175.

\textbf{Perego, J., and S.\ Yuksel} (2022).
``Media Competition and Social Disagreement,''
\textit{Econometrica}, \textbf{90}, 223--265.

\textbf{Prendergast, C.} (1993).
``A Theory of `Yes Men',''
\textit{American Economic Review}, \textbf{83}, 757--770.

\textbf{Prendergast, C., and L.\ Stole} (1996).
``Impetuous Youngsters and Jaded Oldtimers: An Analysis of Behavioral Decision-Making Rules,''
\textit{Journal of Political Economy}, \textbf{104}, 1105--1134.

\textbf{Scharfstein, D., and J.\ Stein} (1990).
``Herd Behavior and Investment,''
\textit{American Economic Review}, \textbf{80}, 465--479.

\textbf{Sharma, M., M. Tong, T. Korbak, D. Duvenaud, A. Askell, S. Bowman, E. Durmus \textit{et al.}} (2024). ``Towards understanding sycophancy in language models.'' \textit{International Conference on Learning Representations}.

\textbf{Sion, M.} (1958).
``On General Minimax Theorems,''
\textit{Pacific Journal of Mathematics}, \textbf{8}, 171--176.

\textbf{Thaler, M.} (2024).
``The Fake News Effect: Experimentally Identifying Motivated Reasoning Using Trust in News,''
\textit{American Economic Journal: Microeconomics}, \textbf{16}(2), 1--38.

\textbf{Thaler, M.} (2025).
``The Supply of Motivated Beliefs,''
CESifo Working Paper 11828.

\appendix
\setstretch{0.98}
\setlength{\parskip}{0.85em}
\section{Benchmark Proofs}

\subsection{Proof of Lemma \ref{lem:1}}\label{sec:prolem1}
Put $e=\bep$ and $C=e-e^2/3$. For $m$ and $x$ in $[0,1]$, define
\[
W(m,x)=\left(1-\frac{|m-x|}{2e}\right)_+
       +\left(1-\frac{m+x}{2e}\right)_+.
\]
Set $A(m)=\int_0^m W(m,x)\de x$. Then
\[
A(m)=\begin{cases}m(2e-m)/e,&m\le e,\\e,&m\ge e.\end{cases}
\]
The required indifference equation is
\begin{equation}\label{eq:max-indifference}
Q(m):=A(m)\phi_{\max}(m)
 +\int_m^1W(m,x)(1-\phi_{\max}(x))\de x=C.
\end{equation}
Symmetry of $W$ gives, for every bounded measurable function $\phi_{\max}$,
\begin{equation}\label{eq:max-average}
\int_0^1 Q(m)\de m
=\int_0^1\int_0^xW(m,x)\de m\de x=C.
\end{equation}

On $[e,1]$, write $\psi=1-\phi_{\max}$ and extend $\psi$ by zero above one.
Equation \eqref{eq:max-indifference} becomes
\[
\psi(m)=\frac e3+\int_m^1 k(x-m)\psi(x)\de x,
\qquad k(t)=\frac1e\left(1-\frac{t}{2e}\right)_+\quad(t\ge0).
\]
Its unique continuous solution is
\[
\psi(m)=\frac e3\left(1+\sum_{n=1}^{\infty}
                  \int_0^{1-m}k^{*n}(t)\de t\right),
\]
where $k^{*n}$ is the $n$-fold convolution on $[0,\infty)$.
The series converges uniformly, since its $n$th integral is at most
$1/(n!e^n)$. Every term is nonnegative, and the sum of the convolution
densities is positive at every $t>0$. Hence $\psi$ is positive and
strictly decreasing whenever $e<1$.
The function $b(m)=e/3+(1-m)/2$ satisfies
\[
b(m)-\frac e3-\int_m^1 k(x-m)b(x)\de x
=\frac{(1-m)(2e-1+m)_+^2}{24e^2}\ge0.
\]
Iteration of the positive integral operator gives
\begin{equation}\label{eq:max-upper-bound}
0<\psi(m)\le\frac e3+\frac{1-m}{2}
\le\frac12-\frac e6<\frac12\qquad(e\le m\le1).
\end{equation}

For $0<m<e$, define
\begin{equation}\label{eq:max-lower-formula}
\phi_{\max}(m)=\frac12+
\frac{1}{4m^2(2e-m)}\int_{2e-m}^{2e+m}
       \{m^2-(x-2e)^2\}\left(\frac12-\psi(x)\right)\de x.
\end{equation}
Only already defined values of $\psi$ enter this expression.
By \eqref{eq:max-upper-bound} and the extension above one,
$0<1/2-\psi(x)\le1/2$. Consequently,
\[
0<\phi_{\max}(m)-\frac12\le\frac{m}{6(2e-m)}\le\frac16.
\]
Thus $\phi_{\max}$ extends continuously to zero with value $1/2$.
For each fixed $x$, the weight
\[
\frac{1}{2e-m}\left(1-\frac{(x-2e)^2}{m^2}\right)_+
\]
increases with $m$ on $(0,e)$, and increases strictly wherever it is positive.
Formula \eqref{eq:max-lower-formula} therefore makes $\phi_{\max}$
strictly increasing on $[0,e)$.

To verify indifference and continuity at $e$, set
\[
J(m)=\frac{1}{2e}\int_{2e-m}^{2e+m}\psi(x)\de x.
\]
Differentiating \eqref{eq:max-lower-formula}, or integrating in the
opposite order, gives
\[
\frac{\de}{\de m}\{m^2(2e-m)\phi_{\max}(m)\}
=m(2e-m)-emJ(m).
\]
For $m<e$ this is equivalent to
\[
A(m)\phi_{\max}'(m)+(4-3m/e)\phi_{\max}(m)=2-m/e-J(m).
\]
Differentiation of $Q$ in \eqref{eq:max-indifference} gives $Q'=0$
by this identity. Hence $Q=C_0$ on $(0,e)$, whereas $Q=C$ on $[e,1]$.
Equation \eqref{eq:max-average} implies $eC_0+(1-e)C=C$, so $C_0=C$.
The integral term in $Q$ has the same limit from either side at $e$.
Since $A(e)=e>0$, the two definitions of $\phi_{\max}(e)$ agree.
This also applies when $e=1$.
We have obtained a continuous, strictly increasing rule with
$\phi_{\max}(0)=1/2$ and $\phi_{\max}(1)=1-e/3$.

Finally, any continuous solution of \eqref{eq:max-indifference} has the
same restriction to $[e,1]$ by uniqueness of the Volterra equation.
On $(0,e)$ the difference between two solutions is
$c/[m^2(2e-m)]$ by the displayed differential equation.
Continuity at zero forces $c=0$. This proves uniqueness.\qed

\subsection{Proof of Lemma \ref{lem:3}}
The benchmark model is a special case of the general environment considered in Theorem \ref{thm:generaltruth}, with $\Gamma=\mathcal U[-1,1]$, $K(\cdot\mid\theta)=\delta_0$, and $\hat{h}(s\mid\theta)
=\frac{1}{2\bar{\epsilon}}\mathbf{1}\{|s-\theta|\leq\bar\epsilon\}$.

Fix $\phi$ in $\Phi^{\mathrm{ind}}$. When the expert reports truthfully and the quack uses $F_Q=\mathcal U[-1,1]$, $\phi$ is Bayes optimal: it selects the unique speaker whose report is consistent with $s$ whenever there is one, while either selection is optimal when both reports are consistent. Conversely, the defining indifference condition for $\Phi^{\mathrm{ind}}$ makes every report optimal for the quack, so $F_Q=\mathcal U[-1,1]$ is a best response. Thus, $(F_Q,\phi)$ is a saddle point of the benchmark truth-telling game. The truth-telling argument in Section \ref{sec:step3} therefore implies that truthful reporting is optimal for the expert for almost every state.

We additionally provide a direct and constructive proof of Lemma \ref{lem:3} in Online Appendix OA.2, which proves optimality for every state and uniqueness under the max and min rules.

\subsection{Proof of Proposition \ref{pr:2}}\label{sec:proppr2}
For the first expression: for each signal realization $s$, the judge learns the state if the quack's message is not consistent, which happens with probability
$$\left(1-\int_{\max\{s-\bar{\epsilon},-1\}}^{\min\{s+\bar{\epsilon},1\}}\frac{1}{2}\de m\right)=\begin{cases}1-\bar{\epsilon}\text{, if }s\in[-(1-\bar{\epsilon}),(1-\bar{\epsilon})],\\
\frac{1-\bar{\epsilon}+s}{2}\text{, if }s\geqslant(1-\bar{\epsilon}),\\
\frac{1-\bar{\epsilon}-s}{2}\text{, if }s\leqslant-(1-\bar{\epsilon}).\end{cases}$$
Combining this with the distribution of the judge's signal, we have
\begin{align*}\mathbf{P}\left\{\hat{G}=\delta_\omega\right\}&=(1-\bar{\epsilon})\int_{-(1-\bar{\epsilon})}^{(1-\bar{\epsilon})}\frac{1}{2}\de s\\
&+2\times\frac{1}{2}\frac{1}{4\bar{\epsilon}}\int_{1-\bar{\epsilon}}^{1+\bar{\epsilon}}(1-(s-\bar{\epsilon}))(1+(s-\bar{\epsilon}))\de s
=1-\bar{\epsilon}+\frac{\bar{\epsilon}^2}{3}.\end{align*}

Now condition on a realized state $\omega=\hat{\omega}$. The judge's signal is uniform on $[\hat{\omega}-\bar{\epsilon},\hat{\omega}+\bar{\epsilon}]$. Relative to the interior value $1-\bar{\epsilon}$, a signal above $1-\bar{\epsilon}$ contributes the excess $(s-1+\bar{\epsilon})/2$, and a signal below $-(1-\bar{\epsilon})$ contributes the excess $(-s-1+\bar{\epsilon})/2$. Integrating these two boundary triangles gives
\[
\mathbf{P}\left(\hat{G}=\delta_\omega\mid\omega=\hat{\omega}\right)=1-\bar{\epsilon}
+\frac{(\hat{\omega}+2\bar{\epsilon}-1)_+^2+(-\hat{\omega}+2\bar{\epsilon}-1)_+^2}{8\bar{\epsilon}}.
\]
This expression is even and nondecreasing in $|\hat{\omega}|$, so more extreme states are weakly more likely to be learned. To establish the final conditional comparison, we show that increasing
the noise raises the relative probability of learning an extreme state
rather than a moderate one. Consider two nonnegative realizations
$0\leq\hat{\omega}'<\hat{\omega}\leq1$. Since $|\omega|$ is uniformly
distributed on $[0,1]$, the ratio of its conditional densities at these
two realizations, conditional on $\{\hat{G}=\delta_\omega\}$, equals
\begin{equation}
\frac{g_{\bar{\epsilon}}(\hat{\omega}|\hat{G}=\delta_\omega)}{g_{\bar{\epsilon}}(\hat{\omega}'|\hat{G}=\delta_\omega)}=\frac{
  \mathbf{P}_{\bar{\epsilon}}\left(\hat{G}=\delta_\omega
    \mid\omega=\hat{\omega}\right)
}{
  \mathbf{P}_{\bar{\epsilon}}\left(\hat{G}=\delta_\omega
    \mid\omega=\hat{\omega}'\right)
}.\label{eq:p1}\end{equation}
We first show that the ratio (\ref{eq:p1}) is nondecreasing in $\bar{\epsilon}$.
To verify this, we calculate
\[
\frac{\partial^2}{\partial\hat{\omega}\,\partial\bar{\epsilon}}
\log\mathbf{P}_{\bar{\epsilon}}\left(\hat{G}=\delta_\omega
  \mid\omega=\hat{\omega}\right).
\]
This cross-derivative is zero when
$0<\hat{\omega}<|2\bar{\epsilon}-1|$. When
$|2\bar{\epsilon}-1|<\hat{\omega}\leqslant1$, it equals
\[
\frac{
  4\left[4\bar{\epsilon}^2
    +(1-\hat{\omega})(3+\hat{\omega}-4\bar{\epsilon})\right]
}{
  \left[(\hat{\omega}+2\bar{\epsilon}-1)^2
    +8\bar{\epsilon}(1-\bar{\epsilon})\right]^2
}>0.
\]

The conditional densities are therefore ordered by the monotone likelihood ratio order, and hence by first-order stochastic dominance. For every cutoff $\bar\omega$ in $(0,1)$, $\mathbf{P}\left(|\omega|>\bar{\omega}
  \mid\hat{G}=\delta_\omega\right)$
is increasing in $\bar{\epsilon}$.

\qed

\subsection{Proof of Proposition \ref{pr:8}}\label{sec:proppr8}
We prove a slightly more general version of the proposition, while retaining the benchmark information structure. Let the state $\omega$ have an arbitrary atomless Borel distribution $G$ on $[-1,1]$, and suppose that the judge observes $s=\omega+\varepsilon$, where $\varepsilon$ is independent of $\omega$ and has a Borel density $h$. Write $v$ for the quack’s value in the corresponding truth-telling game, as introduced in Theorem \ref{thm:generaltruth}. We show that, in any equilibrium and for every estimate $\hat{g}$,
$$\mathbf P[\hat{g}(m_1,m_2,s)=\omega]\leqslant 1-v,$$
and that every equilibrium in which the expert reports truthfully admits an estimate attaining this bound.\footnote{Proposition \ref{pr:8} follows by taking $G=\mathcal U[-1,1]$, $h(e)=\frac{1}{2\bar\epsilon}\mathbf 1\{|e|\le\bar\epsilon\}$, in which case $v=\frac{\bar\epsilon}{2}-\frac{\bar\epsilon^2}{6}$.}

\begin{proof}
Fix an equilibrium of the communication game. Let $F_E^i(\cdot\mid\omega)$ denote the (behavior) messaging strategy of speaker $i$ when he is the expert --- in the equilibria constructed in this paper, $F_E^i$ is truthful, but here it is taken to be arbitrary --- and let $F_Q^i$ denote his messaging strategy when he is the quack. The judge's information at the time of selecting a speaker is $J=(m_1,m_2,s)$.

\medskip\noindent
The joint law of $(\omega,J)$ is
\[
\tfrac12\,G(\de\omega)h(s-\omega)\de s\,\bigl[F^1_E(\de m_1\mid\omega)F_Q^2(\de m_2)+F_Q^1(\de m_1)F_E^2(\de m_2\mid\omega)\bigr].
\]
For $i=1,2$, disintegrate $G(\de\omega)F_E^i(\de m\mid\omega)=M_i(\de m)\gamma_i(\de\omega\mid m)$, where $M_i$ is the marginal law of the message of speaker $i$ when he is the expert, and $\gamma_i(\cdot\mid m)$ is the conditional law of the state given that message.

Conditional on $J$, the posterior law of the state is a mixture of two components, one per identity assignment. Within the component in which speaker $i$ is the expert, the conditional law of $\omega$ is proportional to $h(s-\omega)\gamma_i(\de\omega\mid m_i)$. Since $h(s-\cdot)$ enters as a density factor, an atom of the posterior at a point $\omega'$ requires $\gamma_1(\{\omega'\}\mid m_1)>0$ or $\gamma_2(\{\omega'\}\mid m_2)>0$. Let
\[
A_i(m):=\{\omega':\gamma_i(\{\omega'\}\mid m)>0\},
\]
a countable set for every $m$.\footnote{The maps $(m,\omega')\mapsto\gamma_i(\{\omega'\}\mid m)$ can be chosen jointly measurable, so that the sets below are measurable and the selection rule constructed below is well defined; we omit these routine details.}

For any estimate $\hat{g}$,
\[
\mathbf{P}\left[\hat{g}(J)=\omega\right]=\mathbf{E}\left[\mathbf{P}\left[\omega=\hat{g}(J)\mid J\right]\right],
\]
and the inner conditional probability is positive only if $\hat{g}(J)$ is an atom of the posterior. Hence,
\[
\mathbf{P}\left[\hat{g}(J)=\omega\right]=\mathbf{P}\left[\hat{g}(J)=\omega,\ \omega\in A_1(m_1)\cup A_2(m_2)\right].
\]

\medskip\noindent
Define from $\hat{g}$ the selection rule: select speaker 1 if $\hat{g}(J)$ is in $A_1(m_1)$; select speaker 2 if $\hat{g}(J)$ is in $A_2(m_2)\setminus A_1(m_1)$; select either speaker otherwise. Consider the event that the estimate is correct, yet the selected speaker $i$ is the quack. On this event, $\omega=\hat{g}(J)$ is in $A_i(m_i)$, where $m_i$ is the quack's message. Conditional on the identity assignment and on $m_i$, the state has law $G$, because the quack's message is independent of the state; since $G$ is atomless and $A_i(m_i)$ is countable, the event has probability zero. Therefore,
\[
\mathbf{P}\left[\hat{g}(J)=\omega\right]\leqslant\mathbf{P}\left[\text{the derived rule selects the expert}\right].
\]
In equilibrium, the judge's selection rule is a best response, so her equilibrium payoff is the maximal probability of selecting the expert given the speakers' strategies; the derived rule can do no better. By Proposition \ref{pr:generalbest}, the judge's equilibrium payoff is itself no larger than her payoff in the equilibria with a truth-telling expert, which equals $1-v$. Combining,
\[
\mathbf{P}\left[\hat{g}(J)=\omega\right]\leqslant 1-v.
\]

\medskip\noindent
In an equilibrium in which the expert reports truthfully, select deterministically a speaker with the highest posterior probability of being the expert, choosing speaker 1 in a tie. Let $\hat g$ be that speaker's report. This defines a Borel estimate with the same probability of selecting the expert as the judge's equilibrium rule. This estimate is correct whenever the judge selects the expert; it is also correct in the event that she selects the quack and the quack's report happens to equal the state --- an event of probability zero, since the quack's report is independent of the state and $G$ is atomless. Hence, $\mathbf{P}[\hat{g}(J)=\omega]=\mathbf{P}[\text{the judge selects the expert}]=1-v$, the judge's value in the truth-telling game.
\end{proof}

 \subsection{Proof of Proposition \ref{pr:var}}
For each realization $s$ of the judge's signal, before seeing the messages from the speakers, her belief about the state is that it is uniform over the set $[\max\{-1,s-\bar{\epsilon}\}, \min\{1,s+\bar{\epsilon}\}]$. The variance of this belief (conditional on $s$) is
$$\mathrm{Var}(\omega|s)=\frac{(\min\{1,s+\bar{\epsilon}\}-\max\{-1,s-\bar{\epsilon}\})^2}{12}.$$
Integrating over the realizations of $s$:
$$\mathbf{E}\left[\mathrm{Var}(\omega|s)\right]=2\times\left[\int_0^{(1-\bar{\epsilon})}\frac{1}{2}\frac{\bar{\epsilon}^2}{3}\de s+\int_{(1-\bar{\epsilon})}^{(1+\bar{\epsilon})}\frac{1+\bar{\epsilon}-s}{4\bar{\epsilon}}\frac{(1+\bar{\epsilon}-s)^2}{12}\de s\right]$$
$$=2\times\left[\frac{\bar{\epsilon}^2-\bar{\epsilon}^3}{6}+\frac{\bar{\epsilon}^3}{12}\right]=\frac{2\bar{\epsilon}^2-\bar{\epsilon}^3}{6}.$$

For a given $s$, after seeing the realization of messages $m_E$ and $m_Q$, there are two cases. If only the expert's message is consistent, then the judge learns the truth and the variance of her posterior belief is $0$. If both messages are consistent, then the variance of her posterior belief is $(m_E-m_Q)^2/4$. Therefore
\begin{align*}
&\mathbf{E}\left[\mathrm{Var}(\omega|s,m_1,m_2)|s\right]=\\[1em]
&\int_{\max\{-1,s-\bar{\epsilon}\}}^{\min\{1,s+\bar{\epsilon}\}}\int_{\max\{-1,s-\bar{\epsilon}\}}^{\min\{1,s+\bar{\epsilon}\}}\frac{(m_E-m_Q)^2}{\min\{1,s+\bar{\epsilon}\}-\max\{-1,s-\bar{\epsilon}\}}\frac{1}{8}\de m_Q\de m_E\\[1em]
&=\frac{(\min\{1,s+\bar{\epsilon}\}-\max\{-1,s-\bar{\epsilon}\})^3}{48}.
\end{align*}
Now integrating over $s$, we have
\begin{align*}
&\mathbf{E}\left[\mathbf{E}\left[\mathrm{Var}(\omega|s,m_1,m_2)|s\right]\right]=\\[1em]
&2\times\left[\int_0^{(1-\bar{\epsilon})}\frac{1}{2}\frac{\bar{\epsilon}^3}{6}\de s+\int_{(1-\bar{\epsilon})}^{(1+\bar{\epsilon})}\frac{1+\bar{\epsilon}-s}{4\bar{\epsilon}}\frac{(1+\bar{\epsilon}-s)^3}{48}\de s\right]\\
&=2\times\left[\frac{\bar{\epsilon}^3-\bar{\epsilon}^4}{12}+\frac{\bar{\epsilon}^4}{30}\right]=\frac{5\bar{\epsilon}^3-3\bar{\epsilon}^4}{30}.
\end{align*}
The expected reduction in posterior variance is equal to
$$\frac{2\bar{\epsilon}^2-\bar{\epsilon}^3}{6}-\frac{5\bar{\epsilon}^3-3\bar{\epsilon}^4}{30}=\frac{10\bar{\epsilon}^2-10\bar{\epsilon}^3+3\bar{\epsilon}^4}{30}.$$

Its derivative is
\[
 \bar\epsilon\left(\frac23-\bar\epsilon+\frac25\bar\epsilon^2\right)>0,
\]
because the quadratic in parentheses has negative discriminant.\qed

\section{Extension Proofs}

\subsection{Proof of Proposition \ref{pr:3}}
The proof proceeds in three steps. First, we express the judge’s error probability under any quack reporting distribution in terms of the overlap between the consistency intervals associated with the expert’s and the quack’s reports. Second, a \emph{layer-cake representation}\footnote{For a nonnegative function $f$, the layer-cake representation is $f(x)=\int_0^\infty\mathbf{1}_{\{f(x)>u\}}\de u$, which expresses $f$ as an integral of indicators of its upper-level sets.}, together with Lemma \ref{lem:oa-rearrangement}, reduces the quack’s problem to choosing a centered interval $[-m,m]$ and drawing reports from $G$ conditional on that interval. Finally, we show that the resulting one-dimensional problem has a unique solution $\bar{m}$. The corresponding reporting distribution is therefore maximin; pairing it with a minimax selection rule and invoking Theorem \ref{thm:generaltruth} yields the desired truth-telling equilibrium. The proof uses the following lemma.

\begin{lemma}\label{lem:oa-rearrangement}
Suppose $G$ has a positive density that is symmetric about $0$ and nonincreasing  on $[0,1]$. Define the overlap
\[
 O_{\bep}(m_1,m_2):=\left(1-\frac{|m_1-m_2|}{2\bep}\right)_+.
\]
Among Borel sets $C$ with prescribed $G(C)$,
\[
             \iint_{C\times C}O_{\bep}(m_1,m_2)\,G(\de m_1)G(\de m_2)
\]
is maximized by the centered interval with the same $G$-mass.
\end{lemma}

\begin{proof}
Write $\hat{g}_C=g\mathbf{1}_C$. The triangular kernel is symmetric decreasing
and positive definite: it is $(2\bep)^{-1}$ times the convolution of two
indicators of $[-\bep,\bep]$. Symmetric decreasing rearrangement therefore
weakly raises $\iint O_{\bep}(m_1,m_2)\hat{g}_C(m_1)\hat{g}_C(m_2)\de m_1\de m_2$. Since
$0\le \hat{g}_C\le g$ and $g$ is already symmetric and decreasing, the rearranged
function still lies below $g$.

In the relaxed problem
$0\le \hat{g}\le g$, $\int \hat{g}=G(C)$, a maximizer exists. Its convolution with
$O_{\bep}$ is symmetric decreasing. The bathtub argument\footnote{For fixed total mass and a pointwise upper bound, a linear objective is maximized by filling locations with the highest weights first.} then saturates
the upper bound on a central interval and sets $\hat{g}$ equal to zero
outside it; otherwise moving equal mass from a lower-potential point to a
higher-potential point raises the objective. Thus a maximizer is
$g\mathbf{1}_{[-m,m]}$.
\end{proof}

For fixed speaker strategies, the judge's \emph{Bayes error} is the minimum probability of selecting the quack over all selection rules. It is concave in the pair of quack laws and invariant under exchanging their labels. Averaging a pair with its exchange therefore weakly raises this error. We may thus use a common quack law. A report in a $G$-null set identifies the quack almost surely. Reallocating the singular part of either quack law to $G$ adds nonnegative mass to the overlapping observation laws and cannot reduce the Bayes error. It therefore suffices to consider a common density relative to $G$.

If the quack's strategy has density
$f_Q$ with respect to $G$, the error probability of a
Bayes-optimal judge is
\begin{equation}\label{eq:oa-bayes-uniform-noise}
 \mathcal B(f_Q)=\frac12\iint O_{\bep}(m_1,m_2)\min\{f_Q(m_1),f_Q(m_2)\}
                          \,G(\de m_1)G(\de m_2).
\end{equation}
Indeed, outside the intersection of two consistency intervals, the
judge identifies the expert; on their intersection, whose normalized
length is $O_{\bep}(m_1,m_2)$, the smaller assignment likelihood is the error
likelihood.

For $C_{\bar{f}}=\{f_Q\ge \bar{f}\}$, layer cake gives
\[
 \mathcal B(f_Q)=\frac12\int_0^\infty
 \iint_{C_{\bar{f}}\times C_{\bar{f}}}O_{\bep}\de G\de G\de \bar{f},
 \qquad
 \int_0^\infty G(C_{\bar{f}})\de \bar{f}=1.
\]
Consequently,
\begin{equation}\label{eq:oa-set-ratio}
 \mathcal B(f_Q)\le\frac12\sup_{C:G(C)>0}
 \frac{\iint_{C\times C}O_{\bep}\de G\de G}{G(C)}.
\end{equation}
Equality is attained by $F_Q=G(\cdot|C)$ whenever $C$ maximizes the
ratio. Lemma \ref{lem:oa-rearrangement} then reduces the supremum to
$C=[-m,m]$.

Put $M(m)=G(m)-G(-m)$ and
$J(m)=\int_{-m}^{m}\int_{-m}^{m}O_{\bep}(m_1,m_2)G(\de m_1)G(\de m_2)$.
The ratio $R(m)=J(m)/M(m)$ tends to zero as $m\downarrow0$. If
$m<\bep$, every pair in $[-m,m]^2$ lies within distance $2\bep$. Writing $E(m):=\iint_{[-m,m]^2}|m_1-m_2|\,G(\de m_1)G(\de m_2)$,
direct differentiation gives
\[
 R'(m)=2g(m)\left\{1-
 \frac{2m-E(m)/M(m)^2}{2\bep}\right\}>0.
\]
Hence $R$ is strictly increasing on $(0,\bep)$, so it has no critical point in that interval. At an
interior critical point $m\ge\bep$, differentiation gives
\[
 J(m)=2M(m)\int_{-m}^{m}O_{\bep}(m,m_2)G(\de m_2),
\]
and a second differentiation, with the first-order condition used to
cancel the remaining terms, gives
{\small\[
 R''(m)=\frac{4g(m)}{M(m)}\left[
 g(m)\bigl\{1+(1-m/\bep)_+\bigr\}
 -\frac{G(m)-G(\max\{-m,m-2\bep\})}{2\bep}\right]\le0.
\]}
For $m\ge2\bep$, the fact that $g$ is decreasing gives
$G(m)-G(m-2\bep)\ge2\bep g(m)$. Moreover, for
$\bep\le m<2\bep$, symmetry gives
\[
 G(m)-G(m-2\bep)
 =\int_0^{2\bep-m}g(x)\de x+\int_0^m g(x)\de x
 \ge2\bep g(m).
\]
Equality requires $g$ to be constant on $[0,m]$.
For $m\ge2\bep$, this follows from log-concavity: a nonincreasing concave
log-density that is constant on $[m-2\bep,m]$ is constant on $[0,m]$.
For $\bep\le m<2\bep$, it follows directly from the displayed integral.
If this constant is $c$, then
$R(m)=2c\bep-2c\bep^2/(3m)$ and
$R'(m)=2c\bep^2/(3m^2)>0$.
Equality therefore cannot occur at a critical point.
Consequently, every interior critical point is a strict maximum. Since $R$ starts
increasing, it has at most one such point (otherwise a local minimum would
lie between two strict maxima). Hence $R$ has a unique maximizer
$\bar m$ in $(0,1]$. Equations \eqref{eq:oa-bayes-uniform-noise}--
\eqref{eq:oa-set-ratio} show that the quack law
\[
 \frac{\mathbf 1_{[-\bar m,\bar m]}}{M(\bar m)}\,G
\]
is thus maximin. Pairing it with any minimax selection rule for the judge gives a saddle
point of the truth-telling game. Section \ref{sec:step3} in the proof of Theorem \ref{thm:generaltruth} then verifies the expert's truth-telling incentive, which concludes the construction of the equilibrium in the three-player game.\qed

\subsection{Proof of Proposition \ref{pr:4}}
The proof follows steps analogous to those in the proof of Proposition \ref{pr:3}.

Here, $G=\mathcal U[-1,1]$ and
$p_1>p_2=1-p_1$. As in the preceding proof, reallocating singular report mass to $G$ cannot reduce the Bayes error. Let $f_Q^i$ be the density of quack $i$'s strategy, relative to $G$. The judge’s Bayes risk, that is, the probability that a Bayes-optimal judge selects the quack, is
{\small\begin{equation}\label{eq:oa-asym-risk}
 \mathcal B(f_Q^1,f_Q^2)=\iint O_{\bep}(m_1,m_2)
 \min\{p_2f_Q^1(m_1),p_1f_Q^2(m_2)\}\,G(\de m_1)G(\de m_2).
\end{equation}}

For $\bar{f}\ge0$, define the level sets
$C_1(\bar f):=\{p_2f_Q^1\ge \bar{f}\}$ and
$C_2(\bar f):=\{p_1f_Q^2\ge \bar{f}\}$, and let $c_i(\bar{f})=G(C_i(\bar{f}))$.

The Riesz rearrangement inequality\footnote{Here the inequality states that replacing $C_1$ and $C_2$ by centered intervals of the same lengths weakly increases $\iint_{C_1\times C_2}O_{\bep}(m_1,m_2)\de m_1\de m_2$.} and layer cake imply
{\small\[
 \mathcal B(f_Q^1,f_Q^2)\le\int_0^\infty J_{\bep}(c_1(\bar{f}),c_2(\bar{f}))\de \bar{f},
 \qquad
 \int_0^\infty c_1(\bar{f})\de \bar{f}=p_2,
 \quad \int_0^\infty c_2(\bar{f})\de \bar{f}=p_1,
\]}
where the centered-interval integral has the closed form
\begin{align*}
 J_{\bep}(c_1,c_2)&=\frac12\{H_{\bep}(c_1+c_2)-H_{\bep}(|c_1-c_2|)\},
 \end{align*}
 \begin{align*}
 H_{\bep}(x)&=
 \begin{cases}
 x^2/2-x^3/(12\bep),&0\le x\le2\bep,\\
 \bep x-2\bep^2/3,&x\ge2\bep.
 \end{cases}
\end{align*}
Let $L_{\bep}(x)=J_{\bep}(x,1)$, defined for $x\ge0$. Substitution in the two
pieces of $H_{\bep}$ gives $J_{\bep}(c_1,c_2)\le c_2L_{\bep}(c_1/c_2)$ for $c_2>0$, and $L_{\bep}''(x)\le0$.\footnote{The first inequality is checked on the four regions determined by
$c_1+c_2\lessgtr2\bep$ and $|c_1-c_2|\lessgtr2\bep$; after cancellation it is,
respectively, a nonnegative multiple of $1-c_2$, $(1-c_2)^2$, or
$(1-c_2)^3$. The second follows immediately from
$L_{\bep}''(x)=\{H''_{\bep}(1+x)-H''_{\bep}(|1-x|)\}/2$ (with the sign adjustment
at $x=1$).}
 Jensen's inequality now yields the bound
\begin{equation}\label{eq:oa-asym-bound}
 \mathcal B(f_Q^1,f_Q^2)\le
 p_1L_{\bep}\left(\frac{p_2}{p_1}\right).
\end{equation}

If $p_2/p_1\ge(1-2\bep)_+$, equality is attained by
\[
 f_Q^2\equiv1,\qquad
 f_Q^1(m)=\frac{1}{\bar m}\mathbf 1\{|m|\le\bar m\},\qquad
 \bar m=\frac{p_2}{p_1}.
\]
Indeed, the two weighted densities in the minimum in
\eqref{eq:oa-asym-risk} then have the common height $p_1$ on the smaller
support. If $p_2/p_1<(1-2\bep)_+$, the function $L_{\bep}$ is linear on
$[0,1-2\bep]$. Equality in \eqref{eq:oa-asym-bound} is therefore also
attained with
\[
 \bar m=1-2\bep,\qquad f_Q^2\equiv1,\qquad
 f_Q^1(m)=\bar m^{-1}\mathbf 1\{|m|\le\bar m\}.
\]
Thus $\bar m=\max\{p_2/p_1,(1-2\bep)_+\}$ is one admissible
choice. When $p_2/p_1<1-2\bep$, every
$\bar m$ in $[p_2/p_1,1-2\bep]$ attains the same bound: on its support,
$p_2/\bar m\le p_1$ and consistency has probability $\bep$. A minimax
Bayes rule paired with these maximin densities is a saddle point. Section \ref{sec:step3} in the proof of Theorem \ref{thm:generaltruth} (including its unequal-prior extension) supplies truthful reporting by the expert, which concludes the construction of the equilibrium in the three-player game.\qed

For reference, the judge's error probability is
\begin{equation}\label{eq:oa-asym-error}
 \mathcal E(p_1)=
 \begin{cases}
 p_1L_{\bep}(p_2/p_1),&p_2/p_1\ge(1-2\bep)_+,\\
 p_2\bep,&p_2/p_1<(1-2\bep)_+.
 \end{cases}
\end{equation}
The two expressions agree at the boundary, and differentiation of the
closed form for $L_{\bep}$ shows that $\mathcal E$ is strictly decreasing
in $p_1$. In the second line (which can occur only for $\bep<1/2$), the
favored speaker is selected whenever his report is consistent. Hence the
favored quack earns $\bep$, the disfavored quack earns zero, and the
weighted error is $p_2\bep$.

\section{General Proofs}

\subsection{Proof of Theorem \ref{thm:generaltruth}}
The proof proceeds in three steps. In Section \ref{sec:step1}, we establish the existence of a saddle point: we impose a uniform density cap on the quack’s joint signal-report distribution, then apply Sion's (1958) minimax theorem to each such capped game. We pass to the limit, using regularity of the garbling $K$ to complete the limiting selection rule so that it also deters singular, signal-contingent deviations. Second, in Section \ref{sec:step2}, fixing an arbitrary saddle point, we show that the quack’s equilibrium reports equalize his payoff conditional on almost every signal. Finally, in Section \ref{sec:step3}, Bayes optimality of the judge’s rule allows us to bound the expert’s gain from any misreporting by the corresponding gain to the quack, averaged over the quack signals that generate the same report. The conditional ``equalization'' established in the second step makes this average nonpositive. We therefore establish truth-telling as an optimal strategy for the expert and hence extend the saddle point to a truth-telling equilibrium of the full game.

\subsubsection{Existence}\label{sec:step1}

Define the overlap
\begin{equation*}
 \beta:=\iint\!\int
 \min\{\hat{h}(s\mid \theta),\hat{h}(s\mid \theta')\}\de s\,\Gamma(\de \theta)\Gamma(\de \theta').
\end{equation*}
For fixed $\theta$ and $\theta'$, the inner integral measures the overlap between the judge's conditional signal distributions, and $\beta$ averages this overlap over two independent draws from $\Gamma$. The overlap $\beta$ is strictly positive: Indeed, $L^1(\R)$ is separable, so its set of probability densities can be covered by countably many balls of radius $1/4$. Some inverse image under $\theta\mapsto \hat{h}(\cdot\mid \theta)$ has positive $\Gamma$-mass. For two realizations $\theta$ and $\theta'$ in that set, the $L^1$ distance of their densities is less than $1/2$, and their overlap is $1-|\hat{h}(\cdot\mid \theta)-\hat{h}(\cdot\mid \theta')|_1/2>3/4$. Hence $\beta>0$.

\medskip\noindent
\emph{Capped games.} For $M\ge1$, let
\[
 D_M:=\left\{\mu:\ \mu\ll\lambda\otimes\Gamma,\quad
 0\le\frac{\de\mu}{\de(\lambda\otimes\Gamma)}\le M,\quad
 \mu(\cdot\times\R)=\lambda\right\}.
\]
Thus, $D_M$ is the set of joint distributions of the quack's signal and report induced by strategies whose conditional report densities, relative to $\Gamma$, are bounded above by $M$. This set is convex and compact in the weak topology of $L^1(\lambda\otimes\Gamma)$ by uniform integrability. If
$r_i(t,m)$ is the density of $\mu_i$ in $D_M$, define
\[
 \bar r_i(m\mid \theta):=\int r_i(t,m)K(\de t\mid \theta),
 \qquad 0\le\bar r_i\le M,
\]
the induced density of the quack reporting $m$, conditional on the expert’s signal being $\theta$.
On the judge's observation space, consisting of the ordered pair of reports $(m_1,m_2)$ and the judge’s signal $s$, define the reference probability
\[
 \rho(\de m_1,\de m_2,\de s)
 =\Gamma(\de m_1)\Gamma(\de m_2)
 \frac{\hat{h}(s\mid m_1)+\hat{h}(s\mid m_2)}2\de s.
\]
The judge’s observations have two possible laws, corresponding to the two assignments of speaker roles. Conditional on speaker 1 or speaker 2 being the quack, respectively, the observation laws have $\rho$-densities

\[
 \ell_{1,\mu_1}
 =\frac{2\bar r_1(m_1\mid m_2)\hat{h}(s\mid m_2)}{\hat{h}(s\mid m_1)+\hat{h}(s\mid m_2)},
 \qquad
 \ell_{2,\mu_2}
 =\frac{2\bar r_2(m_2\mid m_1)\hat{h}(s\mid m_1)}{\hat{h}(s\mid m_1)+\hat{h}(s\mid m_2)}.
\]
These densities are defined arbitrarily when the denominator vanishes, and they are each bounded above by $2M$.

The judge's strategy set
$\mathcal J=\{\phi\in L^\infty(\rho):0\le\phi\le1\}$ is weak-star
compact. The payoff is affine, weakly continuous in the two capped
densities, and weak-star continuous in $\phi$. Sion's (1958) minimax theorem
therefore supplies a saddle point $(\mu_{1M},\mu_{2M},\phi_M)$ with value
$v_M$. The values increase with $M$ and satisfy
\begin{equation}\label{eq:oa-value-bounds-general}
 \frac\beta2\le v_M\le\frac12.
\end{equation}
The lower bound follows by having both quacks ignore their signals and
mimic $\Gamma$; the upper bound follows when the judge ignores all data.
Write $v=\lim_Mv_M$.

\medskip\noindent
\emph{Passage to the unrestricted game.} Take $M_n\uparrow\infty$. On
the two-point compactification $\Rbar=[-\infty,+\infty]$, pass to a
subsequence such that
\[
 \mu_{in}\Rightarrow\mu_i^\circ
 \quad\hbox{on }\R\times\Rbar,
 \qquad
 \phi_n\overset{w^*}{\longrightarrow}\phi^\circ
 \quad\hbox{in }L^\infty(\rho).
\]
Fix a Borel version of $\phi^\circ$. Let $O$ be the full-$\lambda$ set in the definition of regularity. For
$t$ in $O$ define the quack’s highest payoff after seeing a signal $t$ (ignoring exceptional reports belonging to a $\Gamma$-null set),
\[
 u_i(t):=\Gamma\text{-}\operatorname*{ess\,sup}_{m\in\R}
 Q_i^{\phi^\circ}(m\mid t).
\]
This function is Borel. Moreover,
\begin{equation}\label{eq:oa-TV-continuity}
 \sup_m|Q_i^{\phi^\circ}(m\mid t)-Q_i^{\phi^\circ}(m\mid t')|
 \le\|\Gamma_t-\Gamma_{t'}\|_{\rm TV},
\end{equation}
so both $u_i$ and $Q_i^{\phi^\circ}(m\mid\cdot)$ are continuous on $O$.

For completeness, the capped saddle inequalities imply
\begin{equation}\label{eq:oa-conditional-value-bound}
 \frac12\int u_1\de\lambda+\frac12\int u_2\de\lambda\le v.
\end{equation}
To see this, fix $\varepsilon>0$ and consider
$S_i=\{(t,m):Q_i^{\phi^\circ}(m\mid t)>u_i(t)-\varepsilon\}$. On the set
of realizations $t$ for which $\Gamma(S_i(t))\ge1/k$, draw $m$ from $\Gamma$
conditioned on $S_i(t)$, and draw from $\Gamma$ otherwise. This test
strategy has density at most $k$ and is feasible in every sufficiently
large capped game. Passing to the weak-star limit, then letting
$k\to\infty$ and $\varepsilon\downarrow0$, gives
\eqref{eq:oa-conditional-value-bound}.

The essential-supremum bound initially has a report exceptional set that
may depend on $t$. Put
\[
 N_i:=\{m:Q_i^{\phi^\circ}(m\mid t)>u_i(t)
          \text{ for some }t\in O\}.
\]
The set is Borel, since the supremum over $t$ may be taken over a
countable dense subset of $O$. It is $\Gamma$-null. Indeed, Fubini says
that for $\Gamma$-almost every $m$ the set of offending $t$ is
$\lambda$-null; by continuity it is relatively open in $O$, and by
regularity every nonempty relatively open subset of $O$ has positive
$\lambda$-mass. The same argument shows that $\Gamma_t\ll\Gamma$ for
every $t$ in $O$: for a $\Gamma$-null set $A$, the continuous function
$t\mapsto\Gamma_t(A)$ vanishes $\lambda$-almost everywhere and hence
everywhere on $O$.

Modify $\phi^\circ$ on the $\rho$-null set involving $N_1$ or $N_2$ so
that a speaker reporting in $N_i$ loses whenever the other report is not
exceptional. Call the completed rule $\phi^*$. For every $t$ in $O$ and
every report $m$,
\begin{equation}\label{eq:oa-all-deviations}
                    Q_i^{\phi^*}(m\mid t)\le u_i(t).
\end{equation}
Thus \eqref{eq:oa-conditional-value-bound} protects the judge against
every quack kernel, including singular deviations correlated with the
quack's signal.

It remains to produce a quack pair that secures $v$. Let $P_i(\mu_i)$
be the observation law under assignment $i$, and let
\[
 B(\mu_1,\mu_2)=\frac12\int\min\{\de P_1(\mu_1),\de P_2(\mu_2)\}
\]
be the Bayes error. For every bounded continuous function on the
compactified observation space, integration against $P_i(\mu_i)$ reduces
to integration against $\mu_i$ of a bounded function continuous on
$O\times\Rbar$, by total variation continuity of $\Gamma_t$ and dominated
convergence in the report. All $\mu_{in}$ and $\mu_i^\circ$ have marginal
$\lambda$, so their weak convergence also holds on this full-measure
subspace. Convergence of these integrals therefore gives
$P_i(\mu_{in})\Rightarrow P_i(\mu_i^\circ)$. Total variation is lower
semicontinuous under weak convergence, so overlap is upper
semicontinuous and
\[
B(\mu_1^\circ,\mu_2^\circ)\ge\limsup_n v_{M_n}=v.
\]
If a limit kernel assigns mass $c_i(t)$ to $\{\pm\infty\}$, redistribute
that mass according to $\Gamma$:
\[
 \widehat F_i(\cdot\mid t)
 =F_i^\circ(\cdot\cap\R\mid t)+c_i(t)\Gamma(\cdot).
\]
Under either assignment the opposing expert report is finite. Escaping
mass therefore contributes no overlap; moving it onto $\R$ adds
nonnegative mass to both finite observation laws and cannot reduce their
pointwise minimum. Hence
$B(\widehat\mu_1,\widehat\mu_2)\ge v$. Together with
\eqref{eq:oa-all-deviations}, this gives, for every quack pair $\mu$ and
every rule $\phi$,
\[
 U(\mu_1,\mu_2,\phi^*)\le v
 \le U(\widehat\mu_1,\widehat\mu_2,\phi).
\]
Thus $(\widehat\mu_1,\widehat\mu_2,\phi^*)$ is a saddle point.

As a remark, if the reduced primitives are symmetric, as in footnote\ \ref{foot:1}, label exchange and simultaneous
reflection of types, reports, signals, and quack signals preserve the
game. The optimal-strategy sets are convex, so averaging a saddle point
over these transformations gives a symmetric saddle point. Positive but
unequal identity priors merely replace the two factors $1/2$ throughout
the capped games and the limiting inequalities.

\subsubsection{Equalization}\label{sec:step2}

Fix any saddle point $(\mu_1^*,\mu_2^*,\phi^*)$ of the truth-telling game, and fix a Borel version of its selection rule.

\begin{lemma}\label{lem:oa-equalization-general}
For each identity $i$, there is a Borel function $u_i:\R\to[0,1]$ such that
\[
 Q_i^{\phi^*}(m\mid t)\le u_i(t)\quad\hbox{for every }m\hbox{, for }\lambda\hbox{-almost every }t,\]
\[
 Q_i^{\phi^*}(m\mid t)=u_i(t)\quad\mu_i^*\hbox{-almost everywhere},
\]
and $\int u_i\de\lambda$ is identity $i$'s quack payoff.
\end{lemma}

\begin{proof}
Define $u_i(t)=\sup_mQ_i^{\phi^*}(m\mid t)$. The supremum is universally
measurable. Applied to $-Q_i^{\phi^*}$, Proposition 7.50 of Bertsekas and Shreve (1978) supplies, for every
$\varepsilon>0$, a universally measurable $m_\varepsilon(t)$ satisfying
$Q_i^{\phi^*}(m_\varepsilon(t)\mid t)\ge u_i(t)-\varepsilon$. After
completion under $\lambda$, take Borel versions. The pure kernel
$t\mapsto\delta_{m_\varepsilon(t)}$ is admissible, so the saddle
inequality gives
\[
 \int(u_i-\varepsilon)\de\lambda
 \le\int Q_i^{\phi^*}(m\mid t)\mu_i^*(\de t,\de m)
 \le\int u_i\de\lambda.
\]
Letting $\varepsilon\downarrow0$ yields equality and therefore
equalization $\mu_i^*$-almost everywhere.
\end{proof}

Let $F_i^*(\cdot\mid t)$ be a Borel disintegration of $\mu_i^*$ with respect to $\lambda$. Let $\eta_i$ denote quack $i$'s marginal distribution of reports. Under the joint law
$$\Gamma(\de\theta)K(\de t\mid\theta)F_i^*(\de m\mid t)$$
of the expert's signal $\theta$, the quack's signal $t$, and the quack's report $m$, let $\tau_i(\de t\mid m)$ denote the conditional law of the quack's signal given his report, so that the report-signal marginal is $\eta_i(\de m)\tau_i(\de t\mid m)$.
Similarly, let $\chi_i(\de\theta\mid m)$ denote the conditional law of the expert's signal given the quack's report.

\begin{lemma}\label{lem:oa-ci-general}
For $\eta_i$-almost every $m$,
\[
 \chi_i(\de \theta\mid m)=\int\Gamma_t(\de \theta)\tau_i(\de t\mid m).
\]
\end{lemma}

\begin{proof}
Conditional on $t$, the quack's report is independent of the expert's
type and that type has law $\Gamma_t$. Conditioning once more on the
report proves the identity.
\end{proof}

\subsubsection{Truth-Telling}\label{sec:step3}

We prove the claim for the identity assignment where speaker 1 is the expert. Let $\kappa_i(\cdot\mid\theta)$ denote the distribution of quack $i$'s equilibrium report conditional on the expert's signal being $\theta$. This distribution is obtained by averaging the quack's equilibrium reporting strategy over the garbled signal $t$:
\[
\kappa_i(A\mid\theta)
:=\int F_i^*(A\mid t)K(\de t\mid\theta)
\]
for every Borel subset $A$ of $\R$.
When speaker 1 is the expert with signal $\theta$, her payoff from report $m$ is
\begin{equation*}
 E_1(m\mid \theta)=\iint q_1(m,m',s)\hat{h}(s\mid \theta)\de s\,\kappa_2(\de m'\mid \theta).
\end{equation*}
The two possible role assignments induce the following laws for the judge's observation $(m,m',s)$ (when speaker $1$ and $2$ are the quack, respectively):
\begin{align*}
\Lambda_1(\de m,\de m',\de s)
&=\kappa_1(\de m\mid m')\Gamma(\de m')
\hat{h}(s\mid m')\de s,\\
\Lambda_2(\de m,\de m',\de s)
&=\Gamma(\de m)\kappa_2(\de m'\mid m)
\hat{h}(s\mid m)\de s.
\end{align*}
(Under $\Lambda_1$, $m'$ is the expert's signal and truthful report; under $\Lambda_2$, $m$ is the expert's signal and truthful report.)

Recall that $\eta_1$ denotes the marginal distribution of speaker 1's report when speaker 1 is the quack.
Let $\sigma=\Gamma+\eta_1$, $a=\de\Gamma/\de\sigma$, and
$b=\de\eta_1/\de\sigma$. Disintegrating over the first report gives
\[
 \Lambda_1=\sigma(\de m)b(m)\xi_m,
  \qquad
 \Lambda_2=\sigma(\de m)a(m)\nu_m,
\]
where $\xi_m(\de m',\de s)=\chi_1(\de m'\mid m)\hat{h}(s\mid m')\de s$ and $\nu_m(\de m',\de s)=\kappa_2(\de m'\mid m)\hat{h}(s\mid m)\de s$.

Bayes optimality of $\phi^*$ implies, for $\sigma$-almost every $m$,
{\footnotesize\begin{equation}\label{eq:oa-bayes-comparisons}
 a(m)\nu_m\le b(m)\xi_m \hbox{ on }\{q_1(m,\cdot,\cdot)<1\},
 \quad
 a(m)\nu_m\ge b(m)\xi_m \hbox{ on }\{q_1(m,\cdot,\cdot)>0\}.
\end{equation}}
Fix a truthful report $\theta$ and an alternative report $\hat m$. Put
$D_{\hat{m}}(m',s)=q_1(\hat{m},m',s)-q_1(\theta,m',s)$. Since $D_{\hat m}>0$ only where $q_1(\theta)<1$
and $D_{\hat{m}}<0$ only where $q_1(\theta)>0$, applying
\eqref{eq:oa-bayes-comparisons} to the positive and negative parts gives
\begin{equation}\label{eq:oa-aggregate-inequality}
 a(\theta)[E_1(\hat{m}\mid \theta)-E_1(\theta\mid \theta)]
 \le b(\theta)\int D_{\hat{m}}\de\xi_\theta.
\end{equation}
By Lemma \ref{lem:oa-ci-general},
\[
 \int D_{\hat{m}}\de\xi_\theta
 =\int[Q_1^{\phi^*}(\hat{m}\mid t)-Q_1^{\phi^*}(\theta\mid t)]
       \tau_1(\de t\mid \theta).
\]
For $\eta_1$-almost every report $m=\theta$, conditional equalization makes the
second term $u_1(t)$ for $\tau_1(\cdot\mid \theta)$-almost every $t$, while
the first is at most $u_1(t)$. The last display is therefore nonpositive,
simultaneously for every $\hat{m}$. If $b(\theta)=0$, the first comparison in
\eqref{eq:oa-bayes-comparisons} gives the same conclusion directly.
Since $a(\theta)>0$ for $\Gamma$-almost every $\theta$, the left side of
\eqref{eq:oa-aggregate-inequality} proves
\[
 E_1(\theta\mid \theta)\ge E_1(\hat{m}\mid \theta)
 \quad\hbox{for every }\hat m
\]
on a set of full $\Gamma$-measure.\\

The argument for identity 2 is identical. Retaining unequal positive identity priors only multiplies the two measures in \eqref{eq:oa-bayes-comparisons} by positive constants and does not change the argument.

The saddle property supplies the quacks' and judge's best responses; Lemma \ref{lem:oa-equalization-general} supplies the quacks' optimality signal by signal. Bayes' rule determines beliefs almost everywhere, and off-path beliefs may be completed to rationalize the prescribed rule. With these beliefs, the saddle point gives a perfect Bayesian equilibrium in which the expert reports truthfully for $\Gamma$-almost every signal. This completes the proof of the theorem.
\qed

\subsection{Proof of Proposition \ref{pr:generalbest}}

Suppose the judge has commitment power, and apply the revelation principle, asking each speaker to report both his information role and his signal. In a direct mechanism, the judge commits to a selection rule as a function of both speakers' reports, as well as her own signal.

Relax the expert's incentive constraint, fixing her at truthful reporting; this can only improve the judge's attainable payoff. For any direct mechanism, consider the selection rule that follows when both speakers claim to be the expert; this rule maps both speakers' signal reports, as well as the judge's private signal, into a probability that each speaker is selected. This rule is a strategy $\phi$ for the judge in the truth-telling game.

One possible strategy for the quack of identity $i$ in $\{1,2\}$ is to claim expertise and choose any report as a function of his signal $t$, that is, any kernel $F_i(\cdot\mid t)$. By the definition of the value in the truth-telling game,
\[
\sup_{F_1,F_2}U(F_1,F_2,\phi)\geqslant v.
\]
Incentive compatibility requires the quack's payoff from reporting his expertise type truthfully to be at least this deviation payoff. That truthful payoff is exactly the probability that the judge selects the quack. Hence every incentive-compatible mechanism makes an error with probability at least $v$, and gives the judge payoff at most $1-v$.

We know from Theorem \ref{thm:generaltruth} that a saddle point of the truth-telling game attains error probability $v$, and that the expert's incentive constraint holds at that saddle point. Thus the upper bound on the judge's payoff is attained at every truth-telling saddle point. With unequal identity priors, average the two quack incentive constraints using their prior probabilities. The same bound and attainment argument apply.
\qed

\end{document}